\documentclass[11pt,a4paper]{emulateapj}
\usepackage{graphicx}
\usepackage{textcomp}
\usepackage{amsmath}
\usepackage{amssymb}
\usepackage{gensymb}
\usepackage{rotating}
\usepackage{epsfig}
\usepackage{amsmath}
\usepackage{longtable}
\usepackage{color}
 \def\gs{\mathrel{\raise0.35ex\hbox{$\scriptstyle >$}\kern-0.6em\lower0.40ex\hbox{{$\scriptstyle \sim$}}}}
 \def\ls{\mathrel{\raise0.35ex\hbox{$\scriptstyle <$}\kern-0.6em\lower0.40ex\hbox{{$\scriptstyle \sim$}}}}

 \def\Msol{\mathrel{\rm M_{\odot}}}
 \def\Lsol{\mathrel{\rm L_{\odot}}}
 \def\Msolyr{\mathrel{\rm M_{\odot}\,yr^{-1}}}
 \def\Wm2{\,\hbox{W}\,\hbox{m}^{-2}}
 \def\gsim{\mathrel{\raise0.35ex\hbox{$\scriptstyle >$}\kern-0.6em\lower0.40ex\hbox{{$\scriptstyle \sim$}}}}
 \def\lsim{\mathrel{\raise0.35ex\hbox{$\scriptstyle <$}\kern-0.6em\lower0.40ex\hbox{{$\scriptstyle \sim$}}}}
 
 \def\pc{\%}

\lefthead{Simpson et al.}  \righthead{S2CLS: An ALMA survey of bright SMGs in S2CLS--UDS}

\begin{document}

\title{The SCUBA-2 Cosmology Legacy Survey: ALMA resolves the
  bright--end of the sub-millimeter number counts}

\author{
J.\,M.\ Simpson,\altaffilmark{1}
Ian Smail,\altaffilmark{1,2}
A.\,M.\ Swinbank,\altaffilmark{1,2}
S.\,C.\ Chapman,\altaffilmark{3}
J.\,E.\ Geach,\altaffilmark{4}
R.\,J.\ Ivison,\altaffilmark{5,6}
A.\,P.\ Thomson,\altaffilmark{1}
I.\ Aretxaga,\altaffilmark{7}
A.\,W.\ Blain,\altaffilmark{8}
W.\,I.\ Cowley,\altaffilmark{2}
Chian-Chou\ Chen,\altaffilmark{1}
K.\,E.\,K.\ Coppin,\altaffilmark{4}
J.\,S.\ Dunlop,\altaffilmark{5}
A.\,C.\ Edge,\altaffilmark{1}
D.\ Farrah,\altaffilmark{9}
E.\ Ibar,\altaffilmark{10}
A.\,Karim,\altaffilmark{11}
K.\,K.\ Knudsen,\altaffilmark{12}
R.\,Meijerink,\altaffilmark{13}
M.\,J.\ Micha{\l}owski,\altaffilmark{5}
D.\ Scott,\altaffilmark{14}
M.\ Spaans,\altaffilmark{15}
P.\,P.\ van der Werf \altaffilmark{13}
}

\setcounter{footnote}{0}
\altaffiltext{1}{Centre for Extragalactic Astronomy, Department of Physics, Durham University, South Road, Durham DH1 3LE, UK; email:  j.m.simpson@dur.ac.uk}
\altaffiltext{2}{Institute for Computational Cosmology, Durham University, South Road, Durham DH1 3LE, UK.}
\altaffiltext{3}{Department of Physics and Atmospheric Science, Dalhousie University, Halifax, NS B3H 3J5 Canada}
\altaffiltext{4}{Centre for Astrophysics Research, Science and Technology Research Institute, University of Hertfordshire, Hatfield AL10 9AB, UK} 
\altaffiltext{5}{Institute for Astronomy, University of Edinburgh,
  Royal Observatory, Blackford HIll, Edinburgh EH9 3HJ, UK}
\altaffiltext{6}{European Southern Observatory, Karl Schwarzschild  Strasse 2, Garching, Germany}
\altaffiltext{7}{Instituto Nacional de Astrof\'isica, \'Optica y Electr\'onica (INAOE), Luis Enrique Erro 1, Sta. Mar\'ia Tonantzintla, Mexico}
\altaffiltext{8}{Department of Physics \& Astronomy, University of Leicester, University Road, Leicester LE1 7RH, UK}
\altaffiltext{9}{Department of Physics, Virginia Tech, Blacksburg, VA 24061, USA }
\altaffiltext{10}{Instituto de F\'isica y Astronom\'ia, Universidad de Valpara\'iso, Avda. Gran Breta\~na 1111, Valpara\'iso, Chile}
\altaffiltext{11}{Argelander-Institute for Astronomy, Bonn University, Auf dem H{\"u}gel 71, D-53121 Bonn, Germany}
\altaffiltext{12}{Department of Earth and Space Sciences, Chalmers University of Technology, Onsala Space Observatory, SE-43992 Onsala, Sweden}
\altaffiltext{13}{Leiden Observatory, Leiden University, P.O. Box 9513, NL-2300 RA Leiden, Netherlands}
\altaffiltext{14}{Department of Physics \& Astronomy, University of
  British Columbia, 6224 Agricultural Road, Vancouver, BC, V6T 1Z1, Canada}
\altaffiltext{15}{Kapteyn Astronomical Institute, University of Groningen, The Netherlands}

\begin{abstract} 
We present high-resolution 870\,$\mu$m ALMA continuum maps of 30
bright sub-millimeter sources in the UKIDSS UDS
field. These sources are selected from deep, 1--degree$^{2}$
850--$\mu$m maps from the SCUBA--2 Cosmology Legacy
Survey, and are representative of the brightest sources in the
field (median $S_{\rm SCUBA\text{--}2}$\,=\,8.7$\pm$\,0.4 \,mJy). We
detect 52 sub-millimeter galaxies (SMGs) 
at $>$\,4\,$\sigma$ significance in our 30 ALMA maps. In 61$^{+19}_{-15}$\,$\pc$\, of the ALMA maps the
single-dish source comprises a blend of $\ge$\,2 SMGs, where
the secondary SMGs are Ultra--Luminous Infrared Galaxies (ULIRGs) with
$L_{\rm  IR}$\,$\gsim$\,10$^{12}$\,$\Lsol$. The brightest SMG
contributes on average 80$^{+6}_{-2}$\,\pc\, of the 
single-dish flux density, and in the ALMA maps containing $\ge$\,2 SMGs
the secondary SMG contributes 25$^{+1}_{-5}$\,\pc\, of the
integrated ALMA flux. We construct source counts and
show that multiplicity boosts the apparent single-dish cumulative
counts by 20\,\pc\, at $S_{870}$\,$>$\,7.5\,mJy, and by
60\,\pc\, at $S_{870}$\,$>$\,12\,mJy. We combine our sample with
previous ALMA studies of fainter SMGs and show that the counts are
well--described by a double power--law with a break at
8.5\,$\pm$\,0.6\,mJy. The break corresponds to a luminosity of
$\sim$\,6\,$\times$\,10$^{12}$\,$\Lsol$ or a star-formation rate of
$\sim$\,10$^{3}$\,$\Msolyr$. For the typical sizes of
these SMGs, which are resolved in our ALMA data with $R_{\rm
  e}$\,=\,1.2\,$\pm$\,0.1\,kpc, this yields a limiting SFR density of
$\sim$100\,$\Msol$\,yr$^{-1}$\,kpc$^{-2}$. Finally, the number density
of $S_{870}$\,$\gsim$\,2\,mJy SMGs is 80\,$\pm$\,30 times higher than that derived from
blank--field counts. An over--abundance of faint SMGs is inconsistent
with line--of--sight projections dominating multiplicity in the
brightest SMGs, and indicates that a significant proportion of these
high--redshift ULIRGs are likely to be physically associated.  

\end{abstract}


\section{Introduction}\label{sec:intro}
The population of dusty galaxies that is detected at sub-millimeter
(sub-mm) wavelengths, SMGs, represent some of the most intense sites
of star formation in the Universe. Sub-mm sources were first
uncovered in surveys with SCUBA at the James Clerk Maxwell 
Telescope (JCMT; e.g.\
\citealt{Smail97,Hughes98,Barger98,Eales99,Pope05,Coppin06}), but subsequently
studied at various facilities (e.g.\
\citealt{Greve04,Greve08,Laurent05,Scott06,Bertoldi07,Weiss09,Austermann10,Lindner11,Aretxaga11}),
and the radio--identified subset of the 
population has been shown to lie at a median redshift of
$z$\,$\sim$\,2.3 \citep{Chapman05}. At these redshifts the typical
flux densities of the sources 
($S_{\nu}$\,$\sim$\,5--15\,mJy) correspond to total infrared
luminosities of $\sim$\,10$^{12}$\,--\,10$^{13}$\,$\Lsol$ (star
formation rates of $\sim$\,10$^{2}$\,--\,10$^3$\,$\Msolyr$; see \citealt{Magnelli12,Swinbank13}),
comparable to local Ultra--Luminous Infrared Galaxies (ULIRGs). The
importance of such prodigious star formation rates (SFRs), and thus
rapid growth in stellar mass at high--redshift has led a number of
authors to suggest that sub-mm sources represent a high--redshift
phase in the evolution of local Elliptical galaxies (e.g.\
\citealt{Lilly99, Genzel03,   Blain04a, Swinbank06b, Tacconi08,
  Hainline11, Hickox12, Toft14,Simpson14}), highlighting their
importance for models of galaxy formation.  

Despite surveys with sub-mm\,/\,mm cameras such as SCUBA--2, LABOCA,
AzTEC, or SPIRE on board {\it Herschel}, uncovering large numbers of
sources, follow--up studies have been hampered by the 
coarse resolution delivered by these single-dish facilities
(typically 15$''$--30$''$ FWHM). At this resolution identifying the
optical\,/\,near--infrared counterparts to the sub-mm emission (i.e.\
resolving the sub-mm source into its constituent SMGs) is
challenging, a problem that is compounded by the expectation that
these heavily dust--obscured galaxies are faint at
optical wavelengths. One route to identifying the SMGs contributing to
each sub-mm source has been to exploit the 
correlation between radio flux density and far--infrared emission in
local galaxies, since 1.4\,GHz imaging with the Very Large Array (VLA)
provides the sub-arcsecond resolution required to pin--point
individual SMGs (e.g.\
\citealt{Ivison98,Ivison02,Ivison04,Ivison07,Smail00,Bertoldi07,Biggs11, 
  Lindner11}). Studies employing this method have successfully
constrained the properties of $\sim$\,50\,\pc\, of the SMG population
\citep{Hodge13}, but they do have limitations: this approach involves
significant assumptions about the multi--wavelength properties of
SMGs,  and it is typically biased towards sources at lower redshift
($z$\,$\lsim$\,2.5) due to the positive K--correction at radio
frequencies.  
 
A further complication to the multi--wavelength identification
procedure is caused by the potential blending of multiple individual
SMGs into a single sub-mm source. Such source blending, or
multiplicity, is somewhat expected given the coarse resolution of
single-dish surveys, but is exacerbated by two further
effects. Firstly, the negative K--correction means that a sub-mm selection
probes a large redshift range ($z$\,$\sim$\,1--8), providing a significant
path length for projection. Secondly, a number of studies have
suggested that the intense star formation in SMGs is predominantly
triggered by merger activity (e.g.\
\citealt{Tacconi08,Engel10,Swinbank10,susie12,Menendez13,Chen15}) and
that the SMG population is strongly clustered
(\citealt{Blain04a,Scott06,Weiss09,Hickox12}, but see also
\citealt{Adelberger05,Williams11}). If SMGs are interacting, or reside in over-densities,
then we may also 
expect to resolve sub-mm sources into physically associated (potentially
interacting) pairs of SMGs. Indeed, studies of sub-mm sources that
employ radio identifications often identify multiple {\it robust}
counterparts to a single sub-mm  source (e.g.\
\citealt{Ivison07}), providing the first indication that multiplicity
is a non--negligible effect.   

Prior to the Atacama Large Millimeter\,/\,sub-millimeter Array
(ALMA), sub-mm interferometry with facilities such as the Plateau de
Bure Interferometer (PdbI) or Sub-Millimeter Array (SMA) offered the
only definitive route to identify the SMGs contributing to single-dish
detected sub-mm sources. However, while these facilities provide the
$\sim$\,1--2$''$ resolution necessary to locate SMGs, their
sensitivity meant that follow--up observations were typically only
possible for a  handful of the brightest sub-mm sources (e.g.\
\citealt{Gear00, Iono06, Wang07, Younger07,Younger09, Dannerbauer08,
  Cowie09, Aravena10, Wang11, Barger12, Smolcic12, Chen13, Ivison13}),
and often at different wavelengths to the initial single-dish
selection.  The first conclusive evidence of multiplicity in sub-mm sources was
presented by \citet{Wang11}, who used observations with the SMA to
show that two bright sub-mm sources were comprised of blends of
two--or--three individual SMGs, with flux densities of
$S_{850}$\,=\,3--5\,mJy, and thought to be at different redshifts. 

Building upon this result, \cite{Barger12} used the SMA to observe 16
SCUBA-detected sources in the GOODS--N field, at $\sim$\,2$''$ resolution. The 
observations resolve three of the sub-mm sources into multiple SMGs,
leading the authors to conclude that $\sim$\,40\,\pc\, of SMGs
brighter than 7\,mJy may be blends of multiple SMGs. However, the SMA
observations have a typical depth of
$\sigma_{860}$\,$\sim$\,0.6--1\,mJy, hence only being sensitive to
secondary SMGs brighter than 3--4\,mJy, and the small number of
sources in the sample leads to significant uncertainties on the
multiplicity fraction. In a similar study, \cite{Smolcic12} showed
that 6\,/\,28 LABOCA 870--\,$\mu$m sources are comprised of blends of
SMG in 1.3--mm follow--up observations with the PdBI, with a further 9
sources not detected. 

%While this again hints at significant
%multiplicity, the results are complicated by the different selection
%wavelengths of the single-dish and interferometric follow--up
%observations.  

The commissioning of ALMA promises a revolution in our understanding
of the SMG population. Indeed, even with the limited capabilities
available in Cycle-0, \citet{Hodge13} obtained robust observations of 88
single-dish sources detected at 870\,$\mu$m in the LABOCA survey of
the Extended {\it Chandra} Deep Field South (LESS). These ALMA,
$1.5''$ resolution, ``snapshot'' observations pin--pointed the SMGs
contributing to the LABOCA sources and showed that at least 35\,\pc\,
of the sources are comprised of $\ge$\,2 SMGs. Furthermore, to recover
the LABOCA flux density, \citet{Hodge13} also showed it is necessary
to include flux in faint sources below their nominal detection
threshold, indicating that a significant proportion of the ALMA maps
contain additional faint 1--2\,mJy SMGs.   

One key result from this ALMA--LESS (ALESS) survey is that despite the sample
containing 12 LABOCA sources above 9\,mJy, only one ALMA--detected SMG is
brighter than this limit. As a result, ~\citet{Karim13} conclude that
due to multiplicity the bright--end of the sub-mm number counts may have
been significantly over--estimated in single-dish surveys, suggesting
a cut-off in the SFR in the most luminous starbursts corresponding to a
potential Eddington limit at 9\,mJy (equivalent to a SFR of
$\sim$\,10$^{3}$\,$\Msolyr$). Although a number of SMGs above this threshold have
been detected in previous interferometric surveys (e.g.\
\citealt{Younger07,Younger09, Barger12,Chen13}).  

To improve the statistics of multiplicity in the brightest sub-mm
sources we have obtained ALMA 870\,$\mu$m follow-up observations of 30
bright (850\,$\mu$m--selected) sub-mm sources in the UKIDSS Ultra Deep
Survey (UDS) field \citep{Lawrence07}. These single-dish targets were
selected from deep, wide--field observations taken as part of the
SCUBA--2 Cosmology Legacy Survey (S2CLS) at the JCMT (14.5$''$ FWHM
resolution), and  are representative of the brightest sources in the
field (Geach et al.\ in prep.). We use the data to measure the
multiplicity in the single-dish population, probe the bright--end of
the number counts and investigate the number density of secondary
SMGs.  

The paper is structured as follows. In \S~2 we discuss our sample
selection, the ALMA observations, and our data reduction. In \S~3 we
describe the construction of our source catalog and provide a
comparison between the ALMA and SCUBA--2 detections. In \S~4 we
discuss the fraction of the single-dish sources that fragment into
multiple SMGs and present the {\it resolved} number counts. Our
conclusions are given in \S~5. We adopt a cosmology with $H_{\rm
  0}$\,=\,71\,km\,s$^{-1}$\,Mpc$^{-1}$, $\Omega_{\Lambda}$\,=\,0.73,
and $\Omega_{\rm m}$\,=\,0.27. Throughout this work error estimates
are from a bootstrap analysis, unless otherwise stated.

%
% Figure1 - ALMA stamps 
\begin{figure*}
 \centering
  \includegraphics[width=0.95\textwidth]{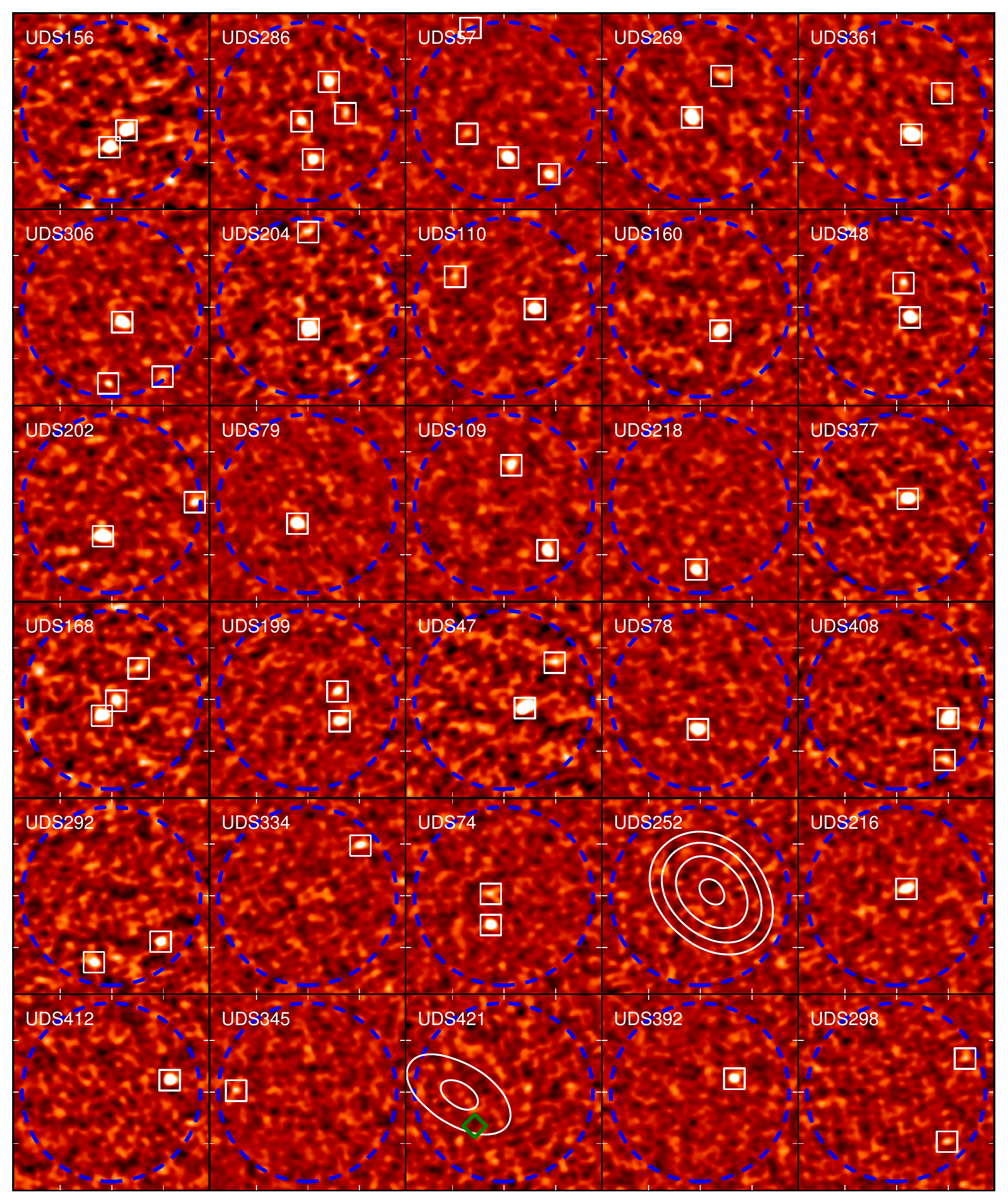}
\caption{ALMA 870--$\mu$m continuum maps, at 0.8$''$ resolution, of 30
  bright sub-mm sources in the UDS field. These sources are selected
  to be representative of the brightest sources detected in the S2CLS
  survey of this $\sim$\,0.8\,deg$^{2}$ field. The
  18$''$\,$\times$\,18$''$ non--primary--beam--corrected maps (roughly
  150\,kpc$\times$\,150\,kpc at the typical SMG redshift, $z$\,=\,2.5)
  are ordered by decreasing single-dish  flux density and have a median 
  1--$\sigma$ rms of 0.26\,mJy\,beam$^{-1}$. The dashed circle on each
  thumbnail represents the primary beam (FWHM) of ALMA at
  870--$\mu$m. We detect 52 SMGs at $>$\,4\,$\sigma$ (marked by a
  squares) in the 30 ALMA maps, with 870--$\mu$m flux densities of
  1.3--12.9\,mJy. In 18\,/\,30 ALMA maps the single-dish sub-mm source
  fragments into two or more individual SMGs. In particular, we
  highlight UDS\,57, 168, 286 and 306, where the ALMA observations
  demonstrate that the single-dish source is comprised of
  three--or--four SMGs. In two ALMA maps, UDS\,252 and 421, we do not
  detect any SMGs, but note that both SCUBA--2 sources are detected in
  {\it Herschel}\,/\,SPIRE imaging. We plot contours representing the
  single-dish SCUBA--2 emission at 3, 3.5, 4.0, 4.5\,$\times \sigma$
  for these sources, note that UDS\,421 has a potential
  VLA\,/\,1.4\,GHz counterpart (diamond; Arumugan et al.\ submitted)
  that is not detected in our ALMA maps.   
}
 \label{fig:stamps}
\end{figure*}

\section{Observations \& Data Reduction}\label{sec:data}
\subsection{Sample Selection}
The sub-mm sources in this paper were selected from observations taken
as part of the S2CLS programme at the JCMT. The latest S2CLS 
map of the UDS field (as of 2014 August) reaches a uniform depth of
$\sigma_{850} = 1.3$\,mJy across 0.78\,deg$^{2}$. However, our initial
sample selection for the Cycle-1 deadline in early 2013 was made from
the first version of these observations (2013 February), which reached a 
1--$\sigma_{850}$ depth of 2.0\,mJy. From these earlier observations we   
selected 31 sources detected at $>$\,4\,$\sigma$, and hence having 
observed 850--$\mu$m flux densities of $>$\,8mJy. We removed one source from
our sample that is a bright, lensed, SMG with previous interferometric
follow--up observations (\citealt{Ikarashi11}), leaving a sample of 30
targets. We note that the sources are extracted from a
 beam--smoothed SCUBA-2 map (i.e.\ matched--filtered), which has a
 resulting spatial resolution of 20.5$''$ (i.e.\ $\sqrt{2}$\,$\times$\,14.5$''$)

\begin{figure}
 \includegraphics[width=0.45\textwidth]{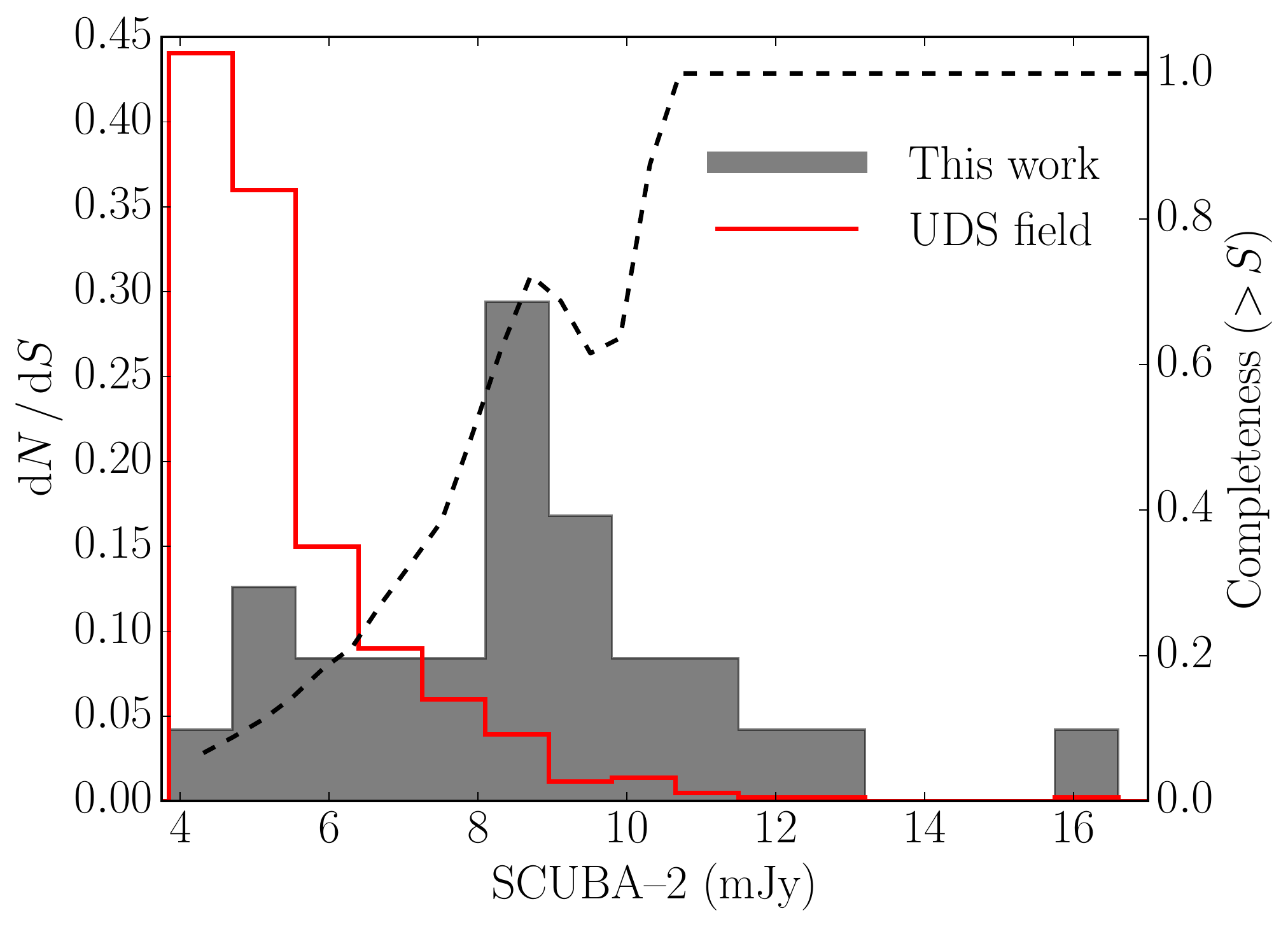} 

\caption{The 850--$\mu$m flux distribution of the
  single-dish--identified sub-mm sources targeted with ALMA (shaded
  histogram) compared to the flux distribution of sub-mm sources in
  the UDS field (open histogram). Both distributions are normalised by
  the total number of sources in each sample. A dashed line shows the
  completeness of our sample, relative to the latest single-dish
  catalog (right--hand axis). While our observations do not represent
  a flux-limited sample in the new S2CLS map, we note that they are
  clearly weighted to the bright-end of the sub-mm population. The
  ALMA sample is $>$50\,\pc\, complete for single-dish sources
  brighter than 8\,mJy, and 100\,\pc\, complete at $>$\,11\,mJy. 
}
 \label{fig:selection}
\end{figure}

In Figure~\ref{fig:selection} we show the flux density distribution for the
30 sources in our sample, measured from the deeper 2014 August S2CLS
map of the UDS field ($\sigma_{850} = 1.3$\,mJy). In the deeper
imaging 12\,/\,30 of the sub-millimeter sources scatter to lower flux
densities ($S_{850}$\,$<$\,$8$\,mJy), with two sources not detected
above our 3.5\,$\sigma$ detection threshold (although both ALMA maps
contain SMGs). While we present the ALMA
observations of these two ``sources'' (UDS\,298 and 392), we note that
formally 28 of the sources in our sample are now
single-dish--detected. Overall our sample consists of
850\,$\mu$m--bright sub-mm sources, with a median observed (i.e.\ not deboosted) flux density of
(8.7\,$\pm$\,0.4)\,mJy. The completeness of our sample relative to the new,
deeper catalog is $>50$\,$\pc$\, at $S_{850}$\,$>$\,$8$\,mJy, and
100\,\pc\, at $S_{850}$\,$>$\,$11$\,mJy over this 0.8\,deg$^{2}$ field.

\subsection{Data Reduction}
We obtained ALMA 870\,$\mu$m (Band 7) continuum imaging of all 30
targets from our sample on 2013 November 1, as part of 
the Cycle-1 project 2012.1.00090.S. All targets were observed using
7.5-GHz of bandwidth centered at 344\,GHz (870\,$\mu$m), chosen to
match the frequency of the original SCUBA--2 observations. We used a
``single continuum'' correlator setup with four basebands of 128
dual-polarization channels each. The FWHM primary beam of ALMA
is 17.3$''$ at our observing frequency, and we centred the
observations at the position of the sub-mm sources in the 2013
February SCUBA--2 map. The ALMA primary beam (FWHM) is
comparable to the spatial resolution of the beam--smoothed SCUBA-2 map
(FWHM\,=\,20.5$''$) and hence our observations are able to detect the
majority of SMGs that contribute significantly to the single-dish source.

The observations were conducted using 26 12--m antennae with a range
of baselines from 20 to 1250\,m, and a median baseline of 200\,m. The
array configuration yields a synthesized beam of 0.35$''$\,$\times
$\,0.25$''$ using Briggs weighting (robust parameter = 0.5), at a
P.A. of $\sim$\,55\,deg for our observations. The observing strategy
involved our 30 targets being observed in two measurement sets, each
containing 15 unique targets. Each measurement set contains seven or
eight sub-blocks, consisting of 30\,s observations of ten targets. In
total each target was observed five times (total integration time of
150\,s), with each repeat distributed randomly within these
sub-blocks. Calibration observations were taken between each
sub-block, with 90\,s phase calibration observations (J\,0217+014; S$_{870}$\,=\,0.49\,Jy) and
30\,s atmospheric calibrations. The absolute flux scale for each
measurement set was derived from observations of J\,0238+166, and
either J\,0423$-$0120 or  J\,0006$-$0623 was used for bandpass
calibration. The flux density of the amplitude calibrator was set at
0.59\,Jy, but we note that the ALMA calibrator archive shows that this
source has day--to--day variations of up to 10\,$\pc$.

The calibration and imaging of our science targets, and calibrators,
was performed using the {\sc 
  Common Astronomy Software Application} ({\sc casa} version
4.2.1).\,\footnote{We repeated the data reduction using the most recent 
version of {\sc casa} (4.2.2) and found it had no effect on our final maps
or source catalog.} To image each target we first
Fourier transform the $uv$--data to create a ``dirty''
map, using Briggs weighting (robust parameter\,=\,0.5). Following
~\citet{Hodge13} we determine the amount of 
cleaning required based on the presence of strong sources in the
maps. We first estimate the RMS in each dirty map and clean the
map to 3\,$\sigma$. We then measure the RMS in the cleaned map and
identify any sources above 5\,$\sigma$. If a source is detected at
$>5$\,$\sigma$ then we repeat the cleaning process but place a tight clean
box around each $>5$\,$\sigma$ source and clean the dirty map to
$1.5$\,$\sigma$. If a map does not contain any
$>5$\,$\sigma$ sources then the map cleaned to 3\,$\sigma$ is
considered the final map. The final maps have a range of
1--$\sigma_{870}$ depths from 0.19--0.24\,mJy\,beam$^{-1}$ (median
$\sigma_{870} = $\,0.21\,mJy\,beam$^{-1}$).

Long wavelength studies to resolve SMGs, either in the
sub-mm, radio,
or molecular line emission (i.e.\ $^{12}$\,CO), suggest that we may
risk resolving the SMGs in our high--resolution ALMA maps
(see~\citealt{Chapman04b,Tacconi06,Biggs08,Younger10,Engel10,Bothwell10,Simpson15}). Hence,
to ensure that any extended flux from the SMGs is not resolved--out,
and that we detect low surface--brightness, extended, sources, we repeated
the imaging process using natural weighting, and applied a 0.6$''$ Gaussian
taper in the $uv$--plane; using a Gaussian taper down--weights
visibilities on long baselines, yielding a larger synthesized beam,
which increases the sensitivity to extended sources, at the cost of
increased noise in the map. The maps were then imaged, and
cleaned, using the same procedure described above to create a set of
low--resolution ``detection'' maps. These low--resolution
``detection'' maps have a median rms of 
$\sigma_{870} = $\,0.26\,mJy\,beam$^{-1}$ and a median synthesized
beam of 0.8\,$'' \times $\,0.65$''$. Both the ``detection'' and
higher resolution maps have a size of 36$''$\,$\times$\,36$''$, and
a pixel scale of 0.04$''$. 

%
% Figure2 Selection

\section{Source Extraction}
To construct a source catalog from our ALMA maps we use the
source extraction package {\sc sextractor}
(v2.8.6;~\citealt{Bertin96}). We search the low--resolution 0.8$''$
``detection'' maps and identify all $>$\,$3$\,$\sigma$ ``peaks'' in 
the non--primary--beam corrected maps. At the position of each
$>$\,$3$\,$\sigma$ detection 
we measure both the peak flux density and the integrated flux density in a
$0.8''$ radius aperture. We also measure integrated flux densities in the
higher resolution maps, at the position of all the detected sources in
the $0.8''$ resolution maps. The fluxes measured in both sets of maps
are primary--beam corrected using the model of the primary beam response
output by {\sc casa}. The integrated flux densities of
  the calibrators, in the 0.8$''$ radius aperture, are 4\,\pc\, lower than
the total flux density measured using the {\sc casa imfit} routine,
and we apply this correction to the integrated flux densities of the
SMGs.

Although we extract sources above $3$\,$\sigma$, we
expect a catalog at this SNR--limit to contain some spurious
detections. To estimate the level of contamination we invert  
the $0.8''$ resolution ``detection'' maps and repeat the source
extraction. Within the FWHM primary beam the number 
of negative detections is lower than positive sources at
$>3.5$\,$\sigma$, but the contamination is $50$\,$\pc$ at
3.5--4.0\,$\sigma$ (falling to $10$\,$\pc$ at 4.0--4.5\,$\sigma$). The
number of sources   detected across all 30 inverted maps is $\le$\,1
at $>$\,$4$\,$\sigma$ (corresponding to a contamination of 2\,\pc\,
when considering our entire catalog) and we therefore adopt this as
the detection threshold for our source catalog. Applying a
$4$\,$\sigma$ cut to our source catalog yields 52 SMGs, within the
FWHM primary beam of the 30 ALMA maps. A search for sources outside
the ALMA primary beam does not identify any statistically significant
detections. We detect no SMGs in two ALMA maps (UDS\,252 and 421) and
a single SMG in a further ten maps. However, in most of the ALMA maps
we detect multiple SMGs and 14, 2, and 2 of the maps contain 2, 3, or 4
SMGs, respectively.  

We perform a number of tests to investigate whether the sources are resolved in the ALMA
imaging. These are detailed in \citet{Simpson15}, but we give a
summary here. First, we measure the ratio of the peak flux in the 0.3$''$ and
0.8$''$ resolution maps. The peak flux of the SMGs is lower in the
0.3$''$ resolution maps, with a median ratio of $S^{0.3}_{\rm
  pk}$\,/\,$S^{0.8}_{\rm pk} = $\,0.65\,$ \pm$\,0.02, indicating that
the sources are resolved in the higher resolution
imaging. Secondly, we investigate the ratio of the integrated--to--peak
flux density in the 0.8$''$ maps; if the sources are unresolved the
peak flux density will equal the integrated flux density. The median
ratio of peak--to--total flux in the 0.8$''$ imaging is 0.82\,$
\pm$\,0.03, again indicating that the sources are marginally resolved
at 0.8$''$ resolution. Finally, we fit point--source and extended
models to the sub-mm emission at both resolutions and find
that a point--source model results in significant residuals, and is
insufficient to describe the emission from these sources. We also show
that the sizes derived from the model fitting are consistent with the
properties of the SMGs in the $uv$--plane \citep{Simpson15}. We
therefore take the flux density of each SMG to be the integrated flux
measured in the 0.8$''$ maps, unless it is lower than the peak flux
density.

%
% Figure 4 - Injected
\begin{figure}
 \includegraphics[width=0.45\textwidth]{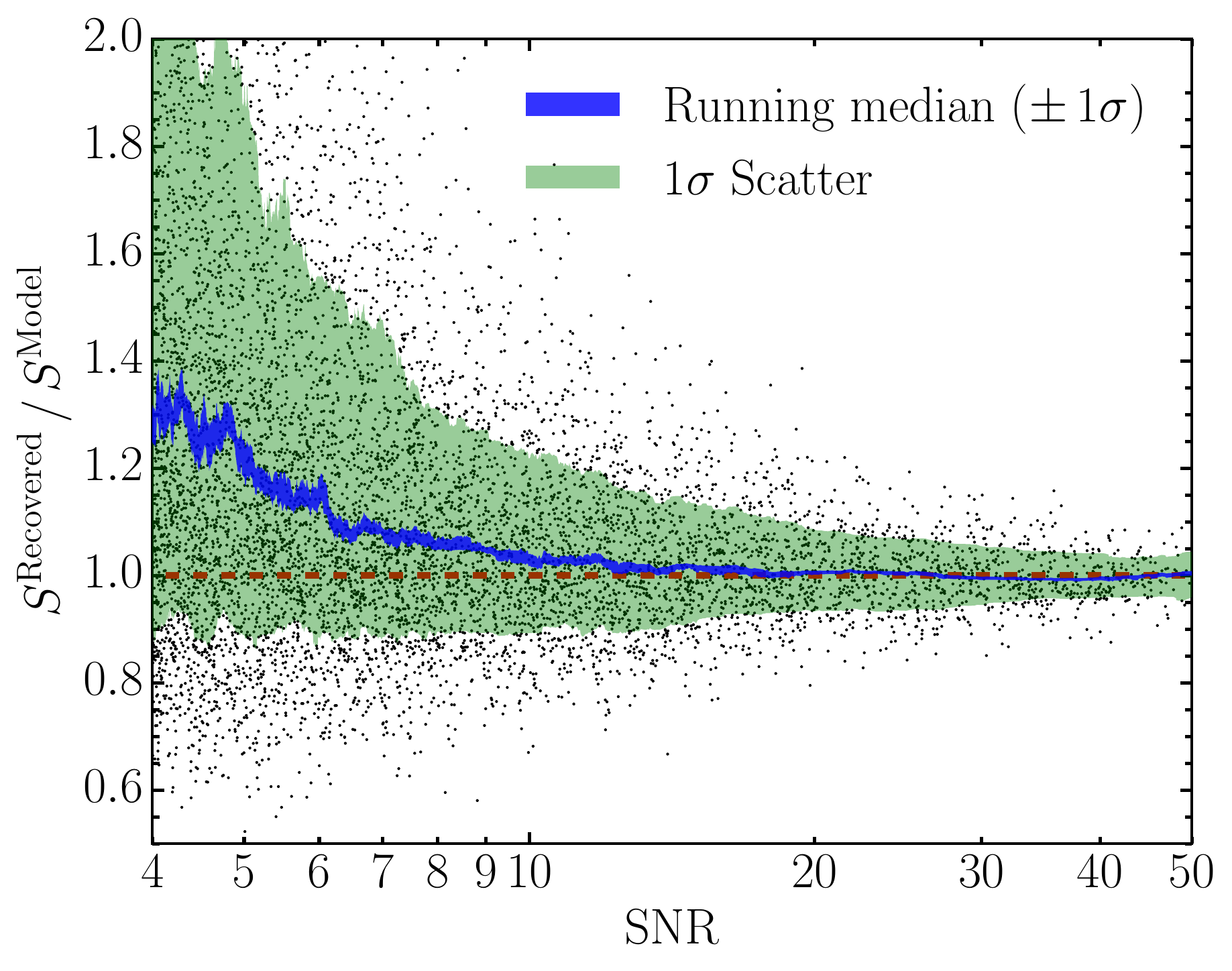} 
 \caption{ The results of simulations involving injecting fake sources
   into our source-subtracted ALMA maps to test the reliability of our
   source extraction procedure. Here we show the ratio of the
   recovered to input source flux density as a function of output
   source SNR, where the flux density of each model source is drawn
  from a steeply declining power-law distribution (with an index of
  $-$2). We show the running median and associated 1--$\sigma$
  bootstrap uncertainty, along-with the 1--$\sigma$ scatter. At our
  detection threshold of 4\,$\sigma$ the flux of individual sources is
  on average boosted by 30\,\pc\,, falling to $<$\,$10$\pc\, at
  $>$\,$6$\,$\sigma$ and $<$\,$1$\pc\, at $>$\,$15$\,$\sigma$.  
}
 \label{fig:injected}
\end{figure}

\subsection{Completeness \& Reliability}
To test the completeness and reliability of our source
extraction we create 2\,$\times$\,10$^{4}$ simulated ALMA
maps. However, to ensure that we have realistic noise properties we
start with one pair (i.e.\ at 0.3$''$ and 0.8$''$ resolution) of our
source--subtracted ALMA maps. To these we then add a model source at
the same, but  random, position in both resolution maps. The flux
densities of the model sources are drawn randomly from a steeply
declining power--law distribution (with an index of $-$2; consistent
with~\citealt{Karim13}), and have peak SNR values of
2--50\,$\sigma$. The intrinsic FWHM size of each model source is drawn from a
uniform distribution from 0--0.5$''$, and we convolve each model
source with the ALMA synthesized beam. To simulate realistic noise, we
add the convolved source to a random position in one pair (i.e.\ at
0.3$''$ and 0.8$''$ resolution) of our source--subtracted ALMA maps.   

%
% Table of properties
%%
 \begin{table*}
 \centering
 \centerline{\sc Table 1: Source Properties}
\vspace{0.1cm}
 {%
 \begin{tabular}{lcccccccc}
 \hline
 \noalign{\smallskip}
ID & R.A.      & Dec.   & $\sigma_{\rm ALMA}$ & $S^{\rm SCUBA\text{--}2}_{\rm obs}$ &
S/N$_{\rm peak}^{\rm ALMA}$   & $S^{\rm ALMA}_{\rm obs}$ & Deboosting$^{b}$ & FWHM$^{c}$   \\
         & (J2000) & (J2000) &  (mJy\,beam$^{-1}$) & (mJy) & & (mJy) & correction &  ($''$)  \\  [0.5ex]  
\hline \\ [-1.9ex]  
  UDS156.0 & 2:18:24.14 & $-$5:22:55.3 & 0.34 & 16.1$\pm$1.2 & 24.5 & 9.7$\pm$0.7 & 1.00 & 0.25$\pm$0.02 \\
  UDS156.1 & 2:18:24.24 & $-$5:22:56.9 & 0.34 & 16.1$\pm$1.2 & 20.0 & 8.5$\pm$0.7 & 1.00 & 0.24$\pm$0.03 \\
  UDS286.0 & 2:17:25.73 & $-$5:25:41.2 & 0.30 & 12.4$\pm$1.2 & 13.3 & 5.2$\pm$0.7 & 0.98 & ... \\
  UDS286.1 & 2:17:25.63 & $-$5:25:33.7 & 0.30 & 12.4$\pm$1.2 & 12.9 & 5.1$\pm$0.6 & 0.98 & 0.26$\pm$0.07 \\
 \hline\hline
 \end{tabular}
 \begin{flushleft}
   \footnotesize{The full version of the catalog is available in the
     online version of this article.  A portion is shown here for form
     and content.\\ $^a$ Source is not detected by SCUBA--2; $^{b}$
     The intrinsic flux densities of the ALMA SMGs are obtained by
     multiplying $S^{\rm ALMA}_{\rm obs}$ with the deboosting
     correction; $^{c}$ Intrinsic source size, corrected for
     synthesised beam (see \citealt{Simpson15})}. Sizes are only
   measured for SMGs detected at $>$\,10\,$\sigma$.
 \end{flushleft}
}
 \refstepcounter{table}\label{table:properties}
 \end{table*}

We perform the source extraction procedure described above on each
simulated map and consider a source recovered if it is detected within 
$0.8''$ of the injected position. The completeness is 93\,\pc\, at
$>$\,4\,$\sigma$, rising to about 100\,\pc\, at 5.5\,$\sigma$,
consistent with the results of similar studies (e.g.\
\citealt{Karim13,Ono14}).  

In Figure~\ref{fig:injected} we show the ratio of output--to--input 
flux density for our simulated sources. The flux densities of sources in a
signal--to--noise limited catalog are known to be boosted if the
sources are drawn from a non-uniform flux distribution. The effect,
known as flux--boosting, arises due to more sources scattering upwards
in flux density, because of random noise fluctuations, than scatter down
as a result of the steeply rising source counts (see also
\citealt{Hogg98,Scott02,Coppin06,Weiss09}). On average a $4$\,$\sigma$
detection in our ALMA maps is boosted in flux by $30$\,$\pc$, with the
boost falling to $<$\,10\,\pc\, at $>$\,6\,$\sigma$ and $<$\,1\,\pc\,
at $>$\,15\,$\sigma$. This flux boosting is sensitive to the slope of
the power-law that defines the flux distribution of the sources, and
we note that varying the slope within the $1$--$\sigma$ uncertainties
presented by~\citet{Karim13} changes the correction by $\pm$\,10\pc\,
at a detection limit of 4\,$\sigma$. We also measure the peak and
integrated flux densities of the simulated sources at both resolutions
and find that the flux boosting correction does not affect our
conclusion that the SMGs in our sample are resolved in the
high--resolution imaging.  

To correct the measured flux densities of the 52 SMGs detected in our
ALMA imaging for flux boosting we calculate the median ratio of 
output--to--input flux for the simulated sources in bins of
0.25\,$\sigma$. We fit a spline to the median of each bin, and
correct the flux densities in our catalog based on the SNR of each
source and the spline fit (Figure~\ref{fig:injected}). Our final
source catalog thus consists of 52 SMGs, with a range of deboosted
870--$\mu$m flux densities of 1.3--12.9\,mJy\,beam$^{-1}$, detected in
30 ALMA maps targeting the brighter single-dish sub-mm sources from
the 0.8\,deg$^{2}$ S2CLS UDS field.  

\subsection{Astrometry and Flux Recovery}
We first compare the flux integrated across all sources in our maps to
that seen by SCUBA--2, to test if our catalog is missing large
numbers of faint sources or if we are missing very extended sub-mm
emission. We also use our catalog to test the astrometry of the
SCUBA--2 map. For each ALMA map we create a model of the
detected SMGs using their primary-beam-corrected flux densities and
convolve each ALMA model map with a model of the SCUBA--2 beam. The
model beam is consistent with a beam created by stacking of bright
SMGs in the UDS field, and with calibrator observations.
 These convolved ALMA maps do not take
into account the contribution to the SCUBA--2 detection from sources
either below the ALMA detection threshold, or outside the primary beam. It also 
neglects any effect due to the  different bandwidth of the SCUBA--2
(35\,GHz half--power bandwidth) and ALMA (2\,$\times$\,4\,GHz)
observations. However, it provides a reasonable test of the effect of a
20\,$\times$ improvement in resolution that our ALMA observations
provide, relative to SCUBA--2.  

\begin{figure*}
 \includegraphics[width=0.45\textwidth]{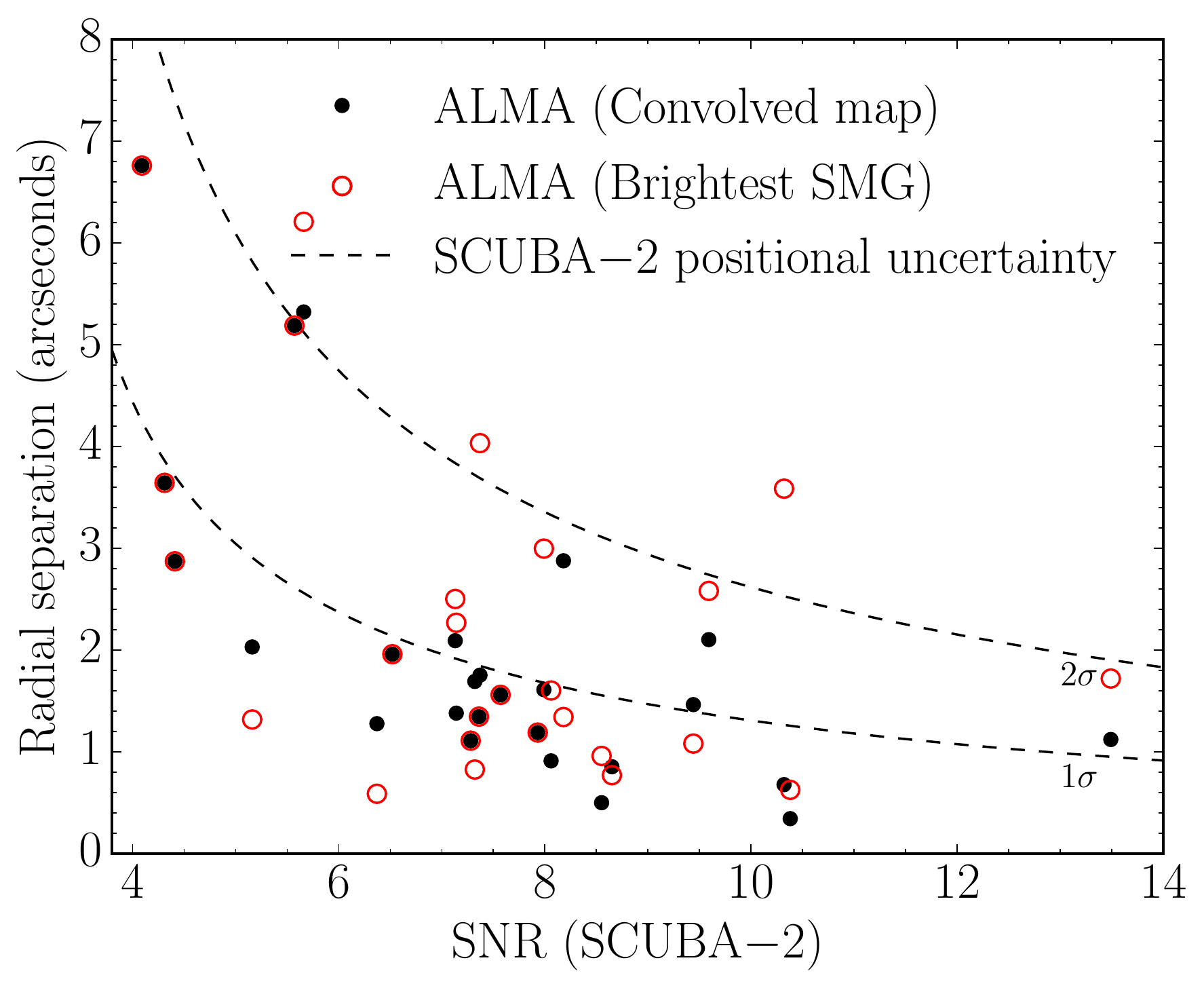} 
 \hfill
 \includegraphics[width=0.45\textwidth]{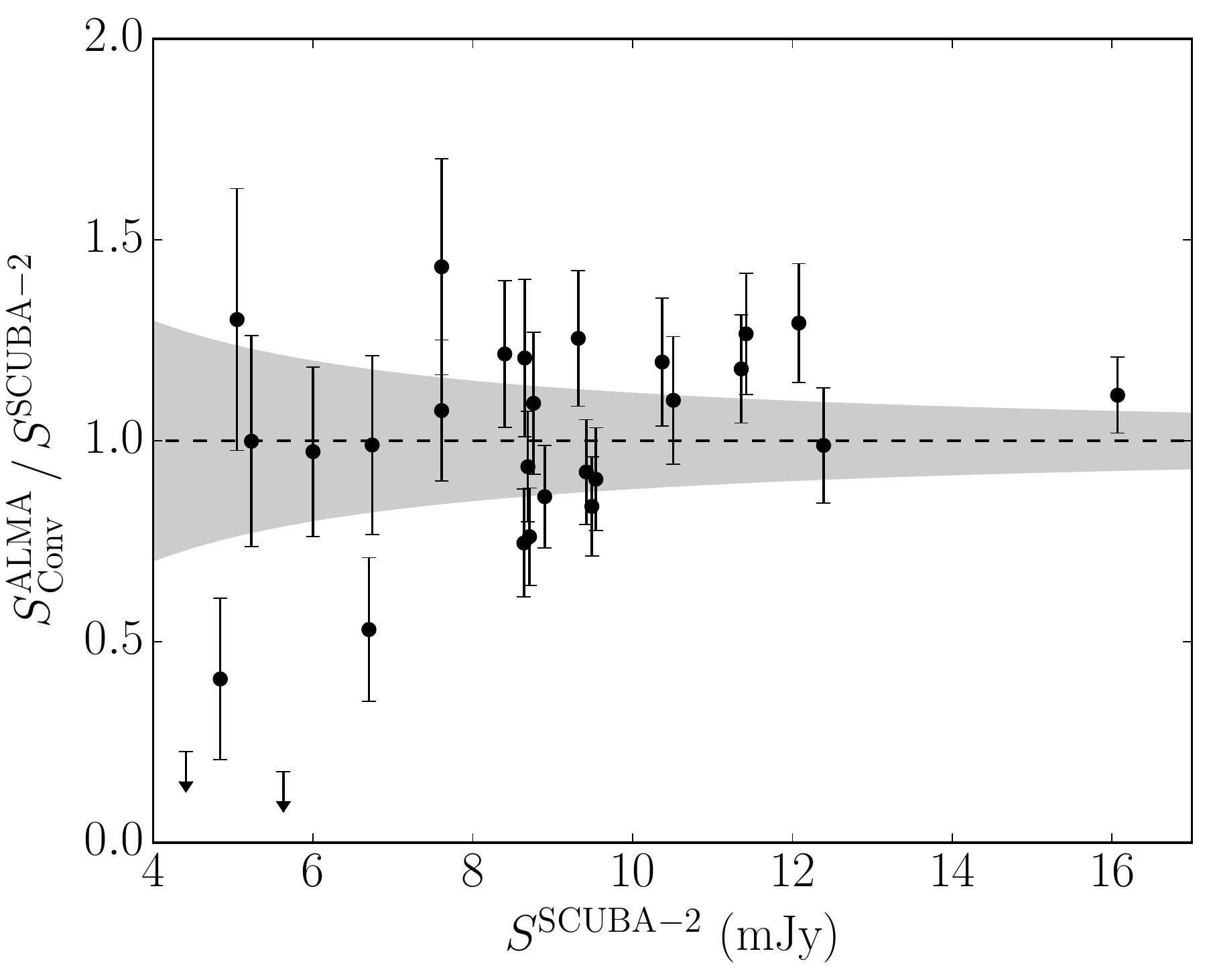} 
\caption{ \textbf{Left:} We convolve the SMGs detected in each ALMA
  map with the SCUBA--2 beam and measure the positional
  offset between the flux centroid in the convolved map and the
  original single-dish-detected sub-millimeter source. The offset
  between the 
  individual ALMA SMGs in each map and the single-dish source is also
  shown. The offsets between the ALMA convolved sources and the
  SCUBA--2 sources are consistent with the predicted uncertainty on the
  SCUBA--2 positions (see \citealt{Ivison07}).  \textbf{Right:} A comparison of
  the peak sub-mm emission from the SMGs detected in each ALMA map,
  convolved with the SCUBA--2 beam, and the SCUBA--2 flux density. The typical
  uncertainty on the SCUBA--2 flux densities is represented by the grey
  shaded region. We find good agreement between the SCUBA--2 and ALMA
  flux densities, with a median ratio of $S^{\rm  ALMA}$\,/\,$S^{\rm
    SCUBA\text{--}2}$\,=\,0.99$^{+0.10}_{-0.04}$. We do not detect any SMGs in
  two ALMA maps, but note that in both cases the 
  SCUBA--2 single dish source is detected in {\it Herschel}\,/\,SPIRE
  imaging at 250, 350, and 500\,$\mu$m indicating that these sources are
  real but potentially faint or multiple SMGs.
}
 \label{fig:offsets}
\end{figure*}

We measure a small, systematic, offset in both R.A.\ and
Dec.\ of $-$0.6$^{+0.3}_{-0.3}$$''$ and $-$1.1$^{+0.2}_{-0.5}$$''$,
respectively (in the sense ALMA$-$SCUBA--2), amounting to less than
the fiducial pixel size of the SCUBA--2 map (2$''$). As expected, the
separation between the SCUBA--2 source and the convolved ALMA map
centroid is a function of the SNR of the single-dish detection
(Figure~\ref{fig:offsets}). Importantly, the measured separations are
consistent with the expected single-dish positional uncertainties:
$70$\,\pc\, of the separations are smaller than the predicted
1--\,$\sigma$ uncertainty on the single-dish position given by
Equation\,B22 from \citet{Ivison07}. These offsets are
between the SCUBA--2 and convolved ALMA peak positions, and hence only
represent the expected search radius for a single, isolated,
counterpart to the single-dish emission (i.e. an ALMA map with a
single detected SMG). However, the median separation between the 
brightest SMG in each map and the SCUBA--2 detection is
1.7$^{+0.6}_{-0.2}$$''$, which is consistent with (although with a
marginally increased scatter) the median separation of the convolved
ALMA map centroids and the SCUBA--2 positions
(1.6$^{+0.2}_{-0.2}$$''$). These results indicate that the offset to
the brightest SMG is consistent with the SNR--based search radius used
to identify counterparts to a sub-mm source, prior to interferometric
observations in the sub-mm (e.g.\ \citealt{Ivison07}).

To confirm the relative flux scales, and also to test that the
observations have not resolved--out flux or missed large numbers of
faint SMGs, we compare the peak flux density of the convolved ALMA
maps to the SCUBA--2 detections (see Figure~\ref{fig:offsets}). The
median ratio  of the ALMA--to--SCUBA--2 flux is $S^{\rm
  ALMA}$\,/\,$S^{\rm   SCUBA\text{--}2}$\,=\,0.99$^{+0.10}_{-0.04}$, including upper limits
for a source at the edge of the primary beam in the ALMA ``blank''
maps. The result indicates good agreement between flux scales, and
suggests that all of the SMGs that contribute significantly to the
single-dish flux density are recovered within the 17.3$''$ ALMA
primary-beam (compared to the 20.5$''$ resolution of the
beam-convolved SCUBA-2 map from which the sources are extracted). We
note that we have not applied a deboosting correction to the SCUBA--2 flux
densities. The deboosting curve for 850\,$\mu$m
SCUBA--2 observations presented by \citet{Chen13b} indicates that the
median deboosting correction for our sample of bright sources is
$\sim$10\,$\pc$. However, we stress that the systematic uncertainty on
the absolute flux calibration of both SCUBA--2 and ALMA are expected
to be comparable, or higher than, the deboosting correction. Indeed,
the systematic uncertainty on the SCUBA-2 flux density scale is
estimated to be 5--10\,\pc\,\citep{Dempsey13}, while the absolute flux
density of the ALMA observations is sensitive to the properties of
the amplitude calibrator chosen. We note that the ALMA data presented
here is calibrated to a Quasar, J\,0238+166, which, as shown in the ALMA 
calibration archive, has daily variations of up to 10\,$\pc$ in
addition to the systematic calibration uncertainties. Given
these large systematic uncertainties we simply note that the flux
density scales of the SCUBA-2 and ALMA data appear 
well--matched with our current analysis.

%
% Figure4 - Positional Offsets

\section{Discussion}
\subsection{Multiplicity}
Previous interferometric follow--up studies of sub-mm sources have
hinted that a fraction of the sources may be comprised of multiple
individual SMGs, which appear blended in the $\gsim$\,15$''$
resolution single-dish imaging
(e.g.\,\citealt{Tacconi06,Ivison07,Wang11,Barger12,
  Smolcic12,Hodge13}). Such an effect is expected, given the low
resolution of single-dish sub-mm maps, but prior to ALMA the effect
has been challenging to quantify due to the small sample sizes, mixed
wavelength of observations and the limited sensitivity of follow--up
studies. 

It is evident from the ALMA maps presented in Figure~\ref{fig:stamps}
that a significant proportion of the single-dish sub-mm sources in
our sample are comprised of multiple, $S_{870}$\,$>$\,1\,mJy,
SMGs. Indeed, 17 of the 28 SCUBA--2 detected sources fragment into
$>$\,1 SMGs, a multiple fraction of $61^{+19}_{-15}$\,\pc\, if we
consider any secondary component, and assuming poisson
uncertainties. In particular we highlight UDS\,57, 168, 286 and 306
where the single-dish sub-mm sources are a blend of three--or--four
SMGs. Hence, each of these maps contains multiple ULIRGs
($S_{870}$\,$\gsim$\,1\,mJy) with a star formation
rate\,\footnote{Assuming a typical conversion between 870--$\mu$m flux
  density and FIR--luminosity (e.g.\ \citealt{Swinbank13}), and a 
  Salpeter Initial Mass Function} of $\gsim$\,150\,$\Msol$\,yr$^{-1}$. 

Defining multiplicity by the number of companions is clearly
dependent on the sensitivity limit for these secondary
components. However, adopting a limit of $S_{870}$\,$>$\,1\,mJy
measures the number of ULIRGs that contribute to each single-dish
sub-mm source, is reasonably well-matched to the depth of our maps and
is sufficiently bright that we would expect to detect $<$\,1 SMG by chance in
our 30 survey fields based on the blank field counts (see
Figure\,\ref{fig:counts}). As we show in \S\,\ref{sec:clustering} the
number density of these secondary SMGs appears to be higher than that
expected in random fields or from simple selection biases,
indicating that a fraction of these multiples are likely to be
physically associated. 

It is interesting to note that two of our ALMA maps are blank, i.e.\ we do 
not detect any SMGs at $>$\,4\,$\sigma$. Although the sub-mm sources
targeted in these maps (UDS\,252 and 421) are two of the fainter SCUBA--2 sources in
our sample (4.4\,mJy and 5.6\,mJy) they are detected in both the 2013
February and 2014 August SCUBA--2 maps, as well as in 250, 350, and 500\,$\mu$m 
{\it Herschel}\,/\,SPIRE imaging (50\,mJy and 26\,mJy at 250\,$\mu$m, respectively),
indicating that they are not simply spurious SCUBA--2 detections. A
simple explanation (given that 17 of our ALMA maps contain multiple
SMGs) is that in these maps the single-dish source is comprised of
multiple SMGs below our detection threshold. In this case,
two--or--three SMGs marginally below the detection threshold at the
edge of the FWHM primary beam ($S_{870}$\,$<$\,2\,mJy) would be
sufficient to explain the missing flux in these maps, and we note that
this would increase the fraction of multiples in our sample to about $70$\,\pc.

\begin{figure*}
 \includegraphics[width=0.45\textwidth]{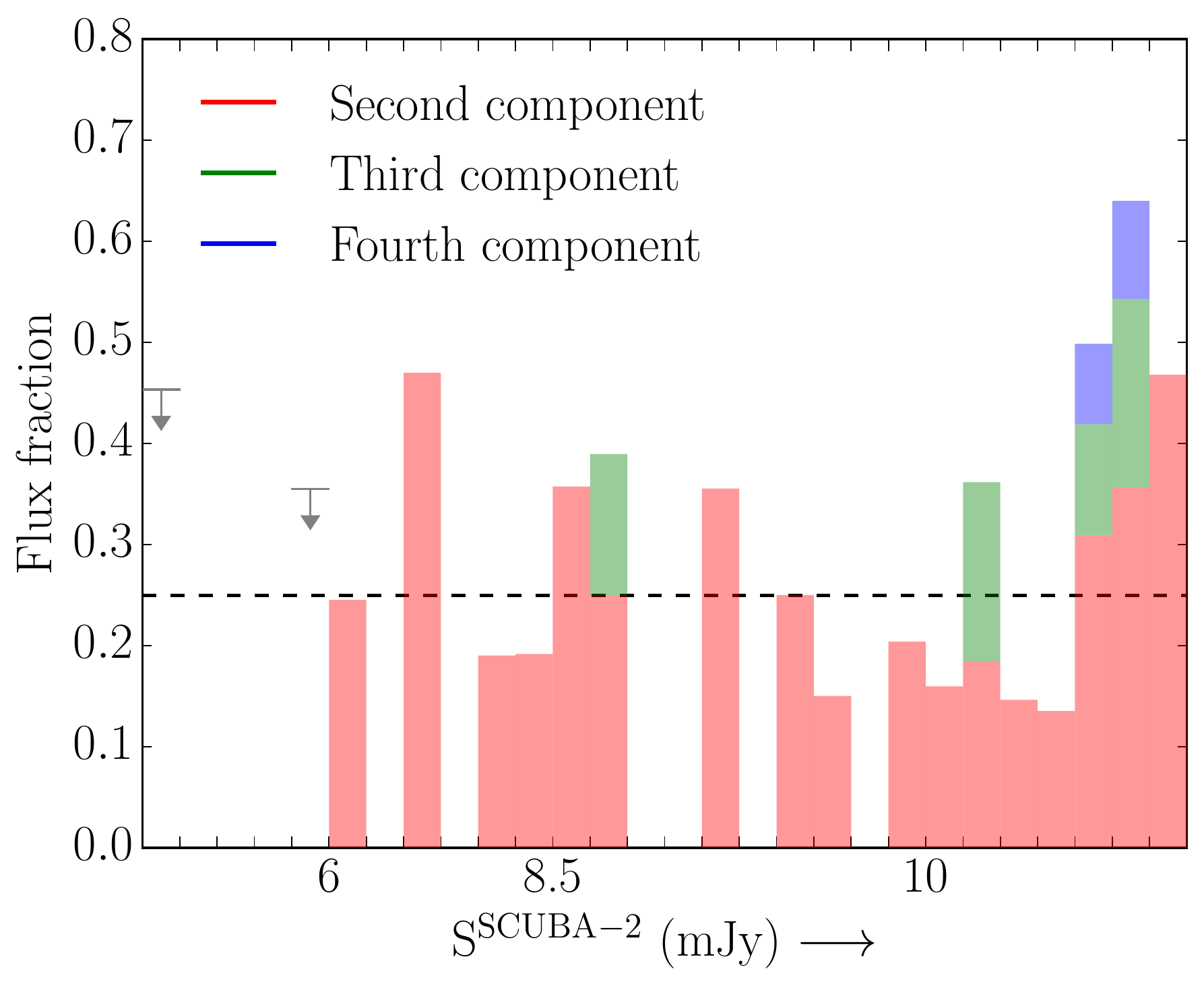} 
 \hfill
 \includegraphics[width=0.45\textwidth]{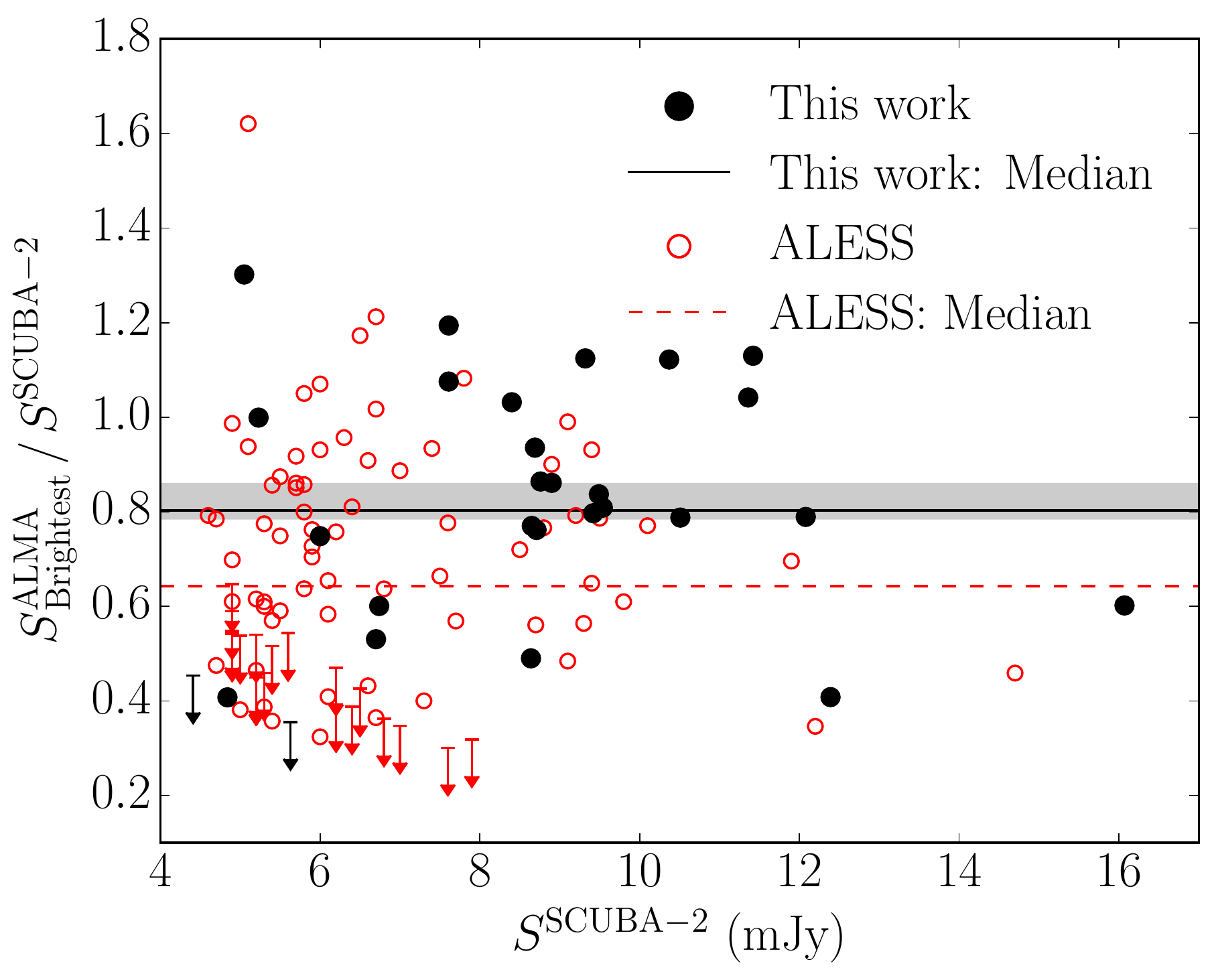} 
\caption{ \textbf{Left:} In 61$^{+19}_{-15}$\,$\pc$\, of our ALMA maps
  the single-dish source targeted comprises of a blend of $\ge$2
  SMGs. Here we show the fraction of the total integrated flux in a
  map that is emitted by each individual SMG. Each interval on the
  abscissa represents an individual ALMA map, and the maps are ordered 
  by increasing single dish flux density. Where an ALMA map contains
  $>$\,1 SMG the second component contributes on average
  25$^{+1}_{-5}$\,\pc\, of the total flux (dashed line), with the third and
  fourth components contributing 16\,\pc\, and 9\,\pc\,,
  respectively. The two ALMA blank maps in our sample are represented
  by upper limits, placed at the maximum that a 4\,$\sigma$
  source at the edge of the ALMA primary beam could contribute to the
  SCUBA--2 flux density. \textbf{Right:} The fraction of the SCUBA--2 flux
  density emitted by the brightest SMG in each ALMA map, as a function
  of single-dish flux density. The median ratio for our sample is
  $S^{\rm ALMA}_{\rm Brightest}$\,/\,$S^{\rm
    SCUBA}$\,=\,0.80$^{+0.06}_{-0.02}$ and
  we do not see a significant trend with single-dish flux
  density. Upper limits correspond to ``blank'' ALMA maps, and are the
  maximum contribution from a $<$\,4\,$\sigma$ source located at the
  edge of the ALMA primary beam. For comparison we show the results
  from the ALESS survey \citep{Hodge13}, which found that 88
  ALMA--identified LABOCA sub-mm sources have a median $S^{\rm}_{\rm
    Brightest}$\,/\,$S^{\rm LABOCA}$\,=\,0.64$^{+0.06}_{-0.03}$. The
  lower fraction of flux density in the brightest component for the
  ALESS sample may be due to the combination of multiplicity and the
  larger beam size of LABOCA (27.2$''$), relative to SCUBA--2 (20.5$''$).  
}
 \label{fig:mult}
\end{figure*}

While the presence of a ULIRG companion to the majority of the
brightest SMGs is clearly significant, it is important to investigate the
relative brightness of the secondary components, and the contribution
they make to the flux density of the original SCUBA--2 detections. In
Figure~\ref{fig:mult} we show the fraction of the 
integrated flux density in an ALMA map that is emitted by each
SMG. We stress that the integrated flux density is the sum of the primary beam
corrected flux densities of the SMGs in each ALMA map, and that this
calculation does not take into account the effect of the SCUBA--2
beam. Where secondary components (i.e.\ fainter SMGs) are detected in
an ALMA map, the ratio between brightest and secondary component is on
average 25$^{+1}_{-5}$\,\pc\,, falling to 16\,\pc\, and 9\,\pc\, for the third and
fourth components, respectively.

As shown in Figure~\ref{fig:mult} we
do not see a significant trend in the fractional flux density of the
secondary components with the single-dish flux density of the
targeted submm sources. To quantify this statement
split the sample into equal subsets at the median single-dish flux
density of the sample ($S_{850}$\,=\,8.7\,mJy). The fainter subset of ALMA maps
have on average 0.5\,$\pm$\,0.2 secondary SMGs, compared to
1.2\,$\pm$\,0.2 for the brighter subset (see
Figure~\ref{fig:mult}). Although the increased number of secondary
SMGs in the ALMA maps of the brightest submm sources tentatively
suggests that brighter single-dish sources are comprised of a blend of
a greater number of SMGs we caution against strong conclusions given
the number of submm sources considered in the analysis. As we note
below this is broadly consistent with the theoretical results of
\citet{Cowley15} who predict that the brightest submm sources are
comprised of a marginally higher number of $S_{870}$\,$\gsim$\,1\,mJy
SMGs.

Next, we investigate the ratio of the brightest component in each ALMA
map to the original SCUBA--2 
detection. We measure a median ratio of $S^{\rm ALMA}_{\rm
  Brightest}$\,/\,$ S^{\rm  SCUBA\text{--}2}$\,=\,0.80$^{+0.06}_{-0.02}$, and
do not find a significant trend in this ratio with single-dish flux
(Figure\,\ref{fig:mult}). This result has important implications
for studies that identify a single counterpart to the sub-mm
emission from a single-dish source using emission at different
wavelengths (e.g.\ 1.4\,GHz); it suggests that even if the
probabilistic identification is correct (see \citealt{Hodge13}) the
true flux density of the SMG is on average $20$\,\pc\, lower than the
single-dish flux. Within the associated errors this
is broadly consistent with the results of~\citet{Cowley15}, who compared
simulated single-dish and interferometric follow--up observations
using the semi--analytic model {\sc galform} and predict that the brightest SMG
comprises $\sim$\,70\,$\pc$ of the single-dish flux density.

We now compare our results to samples of interferometrically
identified SMGs in the literature. \citet{Barger12} present 860--$\mu$m SMA
observations for a sample of 16 850--$\mu$m SCUBA--detected sources and find that
three of the sources are comprised of multiple SMGs. As stated by
those authors, the number of sources in the sample is small, and as the
SMA maps reach a depth $\sigma_{860}$\,$\sim$\,0.7--1\,mJy, they are
only sensitive to secondary SMGs brighter than 3--4\,mJy (at the phase
centre). Similarly, \citet{Smolcic12} found that six out of 28 LABOCA 
sources (870\,$\mu$m selected) fragment into multiple 
components in 1.3--mm observations with the PdBI; in nine of the PdBI
maps no SMGs are detected. While this study again suggests that
multiplicity is important, it is more challenging to interpret as the
single-dish selection and interferometric follow--up observations were
conducted at different wavelengths.   

Recently, \citet{Hodge13} presented the results of an 870--$\mu$m ALMA survey of
122 single-dish sources detected in the 870--$\mu$m LABOCA survey
of the Extended {\it Chandra} Deep Field South (LESS). The reader should note that the
single-dish sub-mm sources studied by \citet{Hodge13} are on average
30\,$\pc$\, fainter than the sources presented here. From the
88 best quality ALESS maps (median $\sigma_{870}$\,=\,0.39\,mJy, with
an interquartile range of 0.35--0.42\,mJy), \citet{Hodge13} extract
a sample of 117 SMGs. In 32 of the maps the single-dish detected
sub-mm source fragments into multiple SMGs. While this indicates that
the fraction of sub-mm sources that are blends of multiple SMGs is
$35$\,\pc, there are two caveats. First, 17 of the
ALMA maps are blank, with the most likely reason being that the
sub-mm source has fragmented into $>$\,$2$ SMGs below the detection
threshold \citep{Hodge13}. Secondly, ~\citet{Hodge13}
show that to recover the LABOCA flux density in the ALMA maps it is
necessary to account for flux from sources below their detection
threshold, indicating that a significant proportion of their ALMA maps
contain faint 1--2\,mJy SMGs, even though sources in this flux range
should be rare in random patches of sky. Support for this conclusion
comes from a stacking analysis of individually undetected IRAC
galaxies in the ALMA maps, which shows that these galaxies are
brighter in the sub--mm than expected for typical IRAC galaxies
\citep{Decarli14}. 

%
% Figure6 - Multiplicity

To perform an accurate comparison between our sample and ALESS we
remove the SMGs from our sample that lie below the ALESS detection
threshold and repeat the multiplicity calculation. In total 13 SMGs
are fainter than the ALESS threshold and are removed from our sample,
resulting in an additional  ``blank'' ALMA map (3\,/\,30). The
fraction of ``blank'' maps in our sample is lower than that for the
ALESS sample, which may simply reflect the significantly lower
resolution of the beam-convolved LABOCA map (27.2$''$ FWHM) compared
to SCUBA-2 (20.5$''$ FWHM).  Of the 17 maps in our 
sample that contain multiple SMGs, seven would have been classed as
single identifications in the ALESS survey and ten as multiples,
yielding a multiplicity fraction of $37^{+15}_{-11}$\,\pc\,. Hence,
the fraction of single-dish sources classed as single identifications
in our survey and ALESS are in close agreement.

To investigate any further differences between the samples we next
compare the fraction of the single-dish flux density in the brightest
component in each ALMA map. The median ratio of the observed flux
densities in the ALESS sample is $S^{\rm ALMA}_{\rm Brightest}
$\,/\,$S^{\rm LABOCA}$\,=\,0.64$^{+0.06}_{-0.03}$, which is lower than
our sample at a 2\,$\sigma$ significance level ($S^{\rm ALMA}_{\rm
  Brightest} $\,/\,$S^{\rm
  SCUBA\text{--}2}$\,=\,0.80$^{+0.06}_{-0.02}$). However, the larger
beam size of LABOCA compared to SCUBA--2 means that secondary
components contribute more to the single-dish detection. To test the
effect of the beam size on the single-dish flux density we convolve a
model of the SMGs in each ALMA map with the SCUBA--2 and LABOCA
beams. We find that on average the LABOCA flux density is $2$\,\pc\,
higher than the SCUBA--2 detection, but stress that this is heavily
weighted by the maps containing a single SMG (where the flux densities
are identical) and that individual sources can be up to $13$\,\pc\,
brighter in the LABOCA observations. While this is clearly a small
effect it does not include sources fainter than the ALMA detection
threshold or outside the ALMA FWHM primary beam and should be
considered a lower limit on the correction. Given all of the results
above, we conclude that the sample presented here and by
\citet{Hodge13} are broadly consistent.

\subsection{Number Counts}
The number counts of SMGs provide one of the most basic
``observables'', which galaxy formation models of the
far--infrared Universe must match. Recently, it has been suggested
that the number of the brightest sub-mm sources
($S_{870}$\,$\gsim$\,9\,mJy) may have been over--estimated in
single-dish studies (e.g.\ \citealt{Karim13}) due to
multiplicity. The single-dish sources in our sample are selected from the central
0.78\,deg$^{2}$ of the S2CLS wide--field map of the UDS and have a median  
flux density of (8.7\,$\pm$\,0.4)\,mJy. The sample is thus ideally suited to
investigate the effects of multiplicity, and measure the intrinsic
form of the bright--end of the number counts.

As discussed earlier our ALMA sample is increasingly incomplete for faint SCUBA--2 sources and so we choose to construct the number
counts from our ALMA source catalog at
$S_{870}$\,$>$\,7.5\,mJy.\,\footnote{The SMGs used to
    construct the number counts are detected at $>$\,15\,$\sigma$ and
    all of the sources would have been detected in our maps, even at the edge of the primary beam.}  To account for the
incompleteness in our sample we first construct the counts from the ALMA observations assuming the
selection is complete. For each ALMA--detected SMG we then 
correct the area surveyed based on the fraction of sources targeted in
the flux bin of the parent single-dish sub-mm source.

\begin{figure*}
 \includegraphics[width=0.45\textwidth]{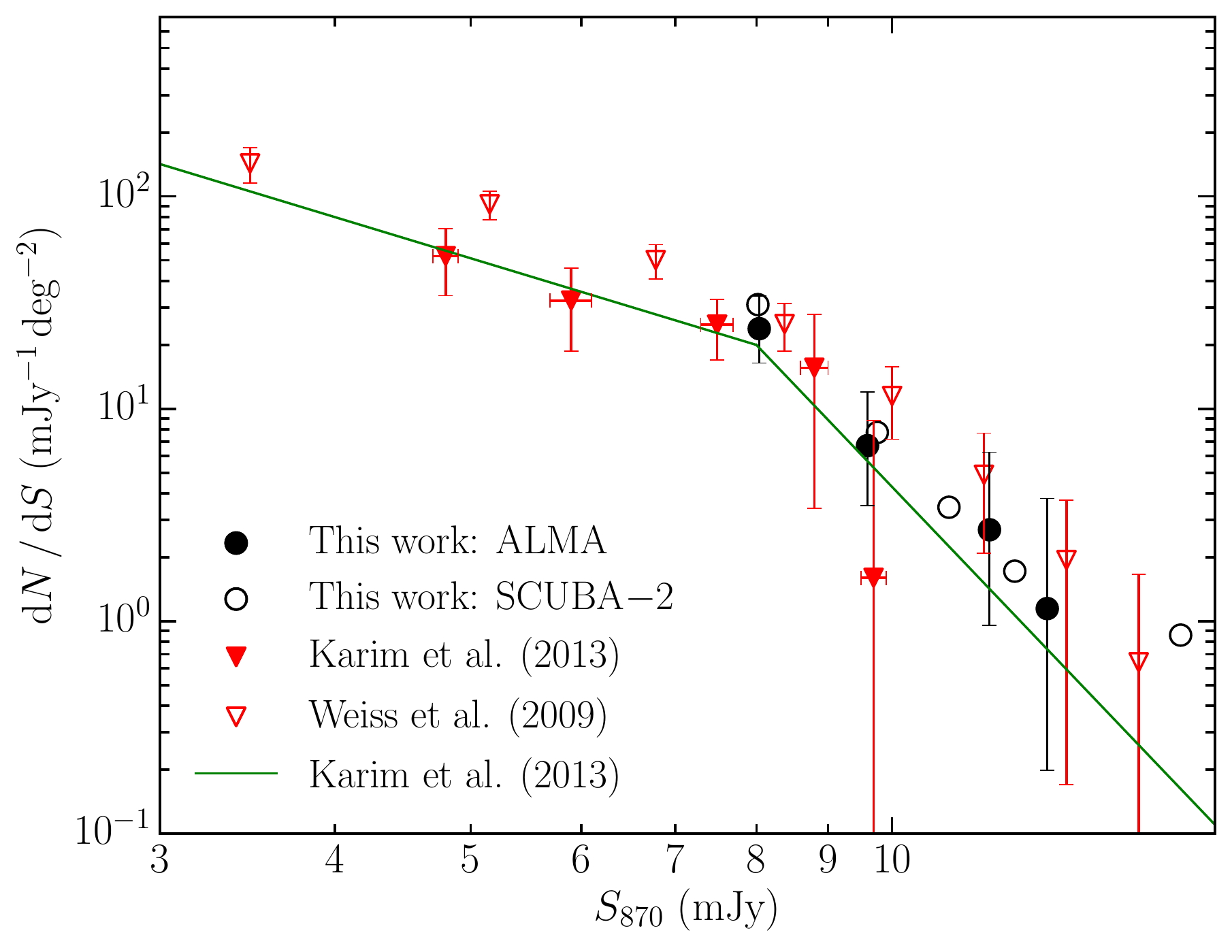} 
 \hfill
 \includegraphics[width=0.45\textwidth]{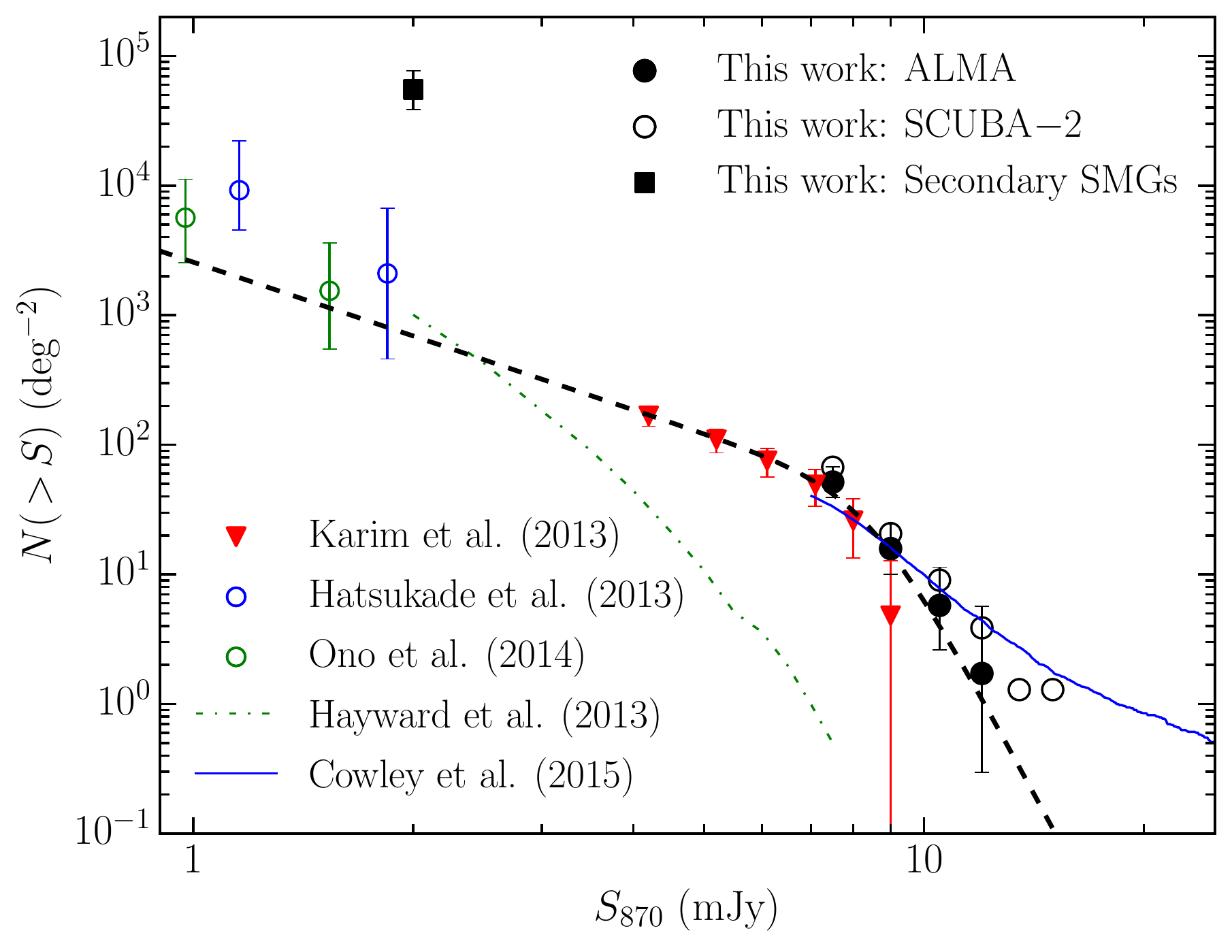}

\caption{ \textbf{Left:} The 870\,$\mu$m differential counts
  constructed from our ALMA observations, compared to the
  parent single-dish SCUBA--2 sample. We also show the counts derived
  from the LABOCA, single-dish survey LESS (\citealt{Weiss09}), and
  the counts from the follow--up ALMA survey ALESS
  (\citealt{Karim13}). The counts derived from our survey are in
  agreement with the ALESS sample at $S_{870}$\,$\lsim$\,$9$\,mJy and
  are well--described by a double--power law (see \S\,4.2). We detect
  seven SMGs with $S_{870}$\,$>$\,$9$\,mJy, compared to one SMG in
  ALESS, and do not see a sharp cut--off in the counts, relative to
  the single-dish observations. \textbf{Right:} Similar to the left
  panel, but instead showing the cumulative counts from the
  870\,$\mu$m surveys. The effect of multiplicity is more obvious in
  the cumulative counts and at $S_{870}$\,$>$\,7.5\,mJy the intrinsic
  counts from our ALMA survey are 20\,\pc\, lower than the counts
  from the parent single-dish sample, falling to 60\,\pc\, lower at
  $S_{870}$\,$>$\,12\,mJy. The cumulative counts from ALMA 
  serendipitous detections at 1--1.3\,mm, converted to 870\,$\mu$m,
  are broadly in agreement with the sample presented here and from
  ALESS. We plot the best--fit double power--law function to all of
  the ALMA samples, which has a break at a characteristic flux density
  of 8.5$^{+0.6}_{-0.6}$\,mJy (dashed line). The theoretical predictions from
  \citet{Cowley15} appear well-matched to the counts presented
  here. However, the counts presented by \citet{Hayward13b} are least
  an order of magnitude lower than the observed counts at
  $S_{870}$\,$>$\,5\,mJy, which is attributed to the absence of
  ``starbursts'' in the model. The number
  density of secondary sources in ALMA maps with a primary SMG
  $>$\,8\,mJy (black square) is a factor of 80\,$\pm$\,30 higher than the
  blank-field counts, which is inconsistent  with these SMGs
  representing a line--of--sight population, and suggests that a
  significant fraction of these SMGs are physically associated.   
}

 \label{fig:counts}
\end{figure*}

In Figure~\ref{fig:counts} we show the differential and cumulative counts
constructed from both our ALMA observations, and the parent SCUBA--2
sample (the uncertainty on the number counts are derived from Poisson
statistics; see \citealt{Gehrels86}). As expected the ALMA number counts
show a decrease relative to the single-dish counts; the
intrinsic cumulative counts are 20\,\pc\, lower than the
single-dish SCUBA--2 counts at $S_{870}$\,$>$\,7.5\,mJy, and
60\,\pc\, lower at $S_{870}$\,$>$\,12\,mJy.

Before discussing the shape and parameterization of the number counts,
we first note that there is a difference between the bright--end of
the number counts presented here, and the ALMA 870--$\mu$m counts
derived from the ALESS survey \citep{Karim13}. Our ALMA observations
targeted eleven single-dish sources with flux densities $>$\,9\,mJy
and detect seven SMGs above this threshold. In contrast,
\citet{Karim13} target 12 single-dish sources brighter than 9\,mJy,
but detect only one ALMA source above this threshold. The difference
between these results may be due to multiplicity and the difference in
the beam sizes of LABOCA and SCUBA--2 (see \S~4.1). However, it is
important to note that the samples are small and are still dominated
by small number statistics.

We now combine our sample with the ALESS
survey \citep{Karim13}, with the aim of providing a single
parameterization of the intrinsic sub-mm number
counts (Figure~\ref{fig:counts}). To extend the range of the number counts to 
lower flux densities we also include two studies that have used serendipitous 
detections of sources in deep targeted ALMA observations to measure the number
counts of faint SMGs at 1.2 and 1.3\,mm from \citet{Ono14} \&
\citet{Hatsukade13}, respectively. Although such studies are sensitive
to clustering between the sources detected serendipitously and the
original targets (which were not selected to be sub-mm sources), they
do provide a crude estimate of the likely number
counts of faint sources. We convert these counts to 870\,$\mu$m
using the composite SMG spectral energy distribution (SED) from the
ALESS survey (see \citealt{Swinbank13}), redshifted to $z$\,=\,2.5
(flux conversion factors are 2.4\,$\times$ and 3.1\,$\times$, at 1.2\,mm and 1.3\,mm,
respectively). Although these converted faint number counts are
sensitive to the shape of the adopted SED, they do appear to be in reasonable
agreement with the cumulative counts from both this study, and ALESS
(see Figure\,\ref{fig:counts}).

%
% Figure8 Counts
 
Since the number counts decline steeply at the bright--end we choose to
model the counts with a double--power law of the form
\begin{equation}
N(>S) = \frac{N_{0}}{S_{0}} \left[ \left(\frac{S}{S_{0}}\right) ^{\alpha} +
\left(\frac{S}{S_{0}}\right)^{\beta} \right]^{-1},
\end{equation}
where $N_{0}$ , $S_{0}$, $\alpha$ and $\beta$ describe the
normalisation, break, and slope of the power laws, respectively. The
best--fit parameters of the model are $N_{0}$\,=\,390$_{-80}^{+110}$\,deg$^{-2}$,
$S_{0}$\,=\,8.4$_{-0.6}^{+0.6}$\,mJy, $\alpha$\,=\,1.9$_{-0.2}^{+0.2}$ and
$\beta$\,=\,10.5$_{-3.2}^{+3.0}$, and as can be seen in
Figure~\ref{fig:counts} the parameterization provides an adequate
representation of the cumulative counts. However we caution that the
number of sources at both the bright and faint end of the counts
remains low (26 and 20 at $S_{870}$\,$<$\,2\,mJy and
$S_{870}$\,$>$\,8\,mJy, respectively) and this is reflected in the
uncertainties on the best-fit parameters.  

When constructing the observed sub-mm number counts we have included
three SMGs from our sample that we identify as potential gravitationally
lensed sources (UDS\,109.0, UDS\,160.0, and UDS\,269.0). All three of these
SMGs appear to be close to, but spatially offset from, galaxies at
$z\ls$\,1 (see \citealt{Simpson15}). Although there are no indications
that these SMGs are strongly lensed (i.e.\ no multiple images), even a
modest magnification of $\mu$\,$\ge$\,1.7 is sufficient to push the
intrinsic flux density of these sources below our threshold for
constructing the number counts. If we remove these three SMGs then the
cumulative number counts decrease by 18\,\pc\, at
$S_{870}$\,$>$\,7.5\,mJy and the parameters of the best--fit double
power--law, for all of the ALMA samples, change by less than their
associated 1--$\sigma$ uncertainties. 

It has been suggested that an absence of bright SMGs
($S_{870}$\,$\gsim$\,9\,mJy) may indicate a physical limit to the
intense starbursts that are occurring in these sources (see
\citealt{Karim13}). We detect bright SMGs in our survey and do not
find evidence for a sharp cut--off in the counts. However, we do find
that number counts decline strongly towards the bright-end with a
distinct break at a flux density of 
$S_{0}$\,=\,8.4$_{-0.6}^{+0.6}$\,mJy, which may suggest a typical
threshold to the SFR. If we adopt the relationship
between $S_{870}$ and $L_{\rm FIR}$ for the ALESS SMGs
\citep{Swinbank13}, and the SMGs presented here (Ma.\ et al.\ in
prep.), then this break corresponds to a luminosity of
$\sim$\,6\,$\times$\,10$^{12}$\,$\Lsol$, or a SFR of
$\sim$10$^{3}$\,$\Msolyr$ (for a Salpeter IMF). The SMGs in our sample
are resolved in our ALMA imaging, and the brightest SMGs have a median
half-light radius of 1.2\,$\pm$\,0.1\,kpc \citep{Simpson15}. Given the
sizes of the SMGs, the break in the number counts corresponds to a
typical threshold to the star formation rate density in these starbursts of
$\sim$\,100\,$\Msol$\,yr$^{-1}$\,kpc$^{-2}$. The star formation rate
density of a typical SMG at the break in the number counts is an order
of magnitude lower than the expected Eddington limit for these sources
(see \citealt{Andrews11,Simpson15}). However, we stress that the star
formation rates are integrated across the whole star forming
region. If the star formation in these SMGs is occurring in individual
``clumps'' (e.g.\ \citealt{Swinbank11,Danielson11,Danielson13}  then
these individual regions may be Eddington limited, while the overall
star forming region appears sub-Eddington.

\subsection{Comparison to galaxy formation models}\label{sec:galform}
We now compare our results to recent theoretical predictions for
sub-mm number counts, which attempt to simulate the effects of blending
in single-dish surveys. \citet{Hayward13b} construct single-dish and
intrinsic sub-mm number counts, based on a hybrid numerical model. By
construction the 
single-dish counts from the model are in broad agreement with
single-dish observations at $S_{870}$\,$\sim$\,5\,mJy but, as shown
in Figure~\ref{fig:counts}, the intrinsic cumulative number
counts under--predict the observed counts by over an order of
magnitude at $S_{870}$\,$>$\,5\,mJy. As stated by \citet{Hayward13b}
the deficit is likely due to the absence of merger--driven
``starbursts'' in the model that act to elevate the star formation
in these systems. We note that a previous model that includes ``starbursts''
is in closer agreement with the observed counts (see
\citealt{Hayward13a}). However, that model has a limited treatment of
source blending and as shown in \citet{Hayward13b} multiplicity has an
order of magnitude effect on their predictions, which is not seen in
our data.

%
% Table of properties
%%
 \begin{table}
 \centering
 \centerline{\sc Table 2: Cumulative 870$\mu${\rm m} number counts}
\vspace{0.1cm}
 {%
 \begin{tabular}{lcccccccc}
 \hline
 \noalign{\smallskip}
 $S_{870}$ & $N$\,(\,$>$\,$S_{870}$)   \\
     (mJy)           &    (deg$^{-2}$)    \\  [0.5ex]  
\hline \\ [-1.9ex]  
\vspace{0.5mm}
 7.5 & 51.6$^{+15.8}_{-12.4}$ \\
\vspace{0.5mm}
 9.0 & 15.8$^{+8.5}_{-5.8}$ \\
\vspace{0.5mm}
 10.5 & 5.8$^{+5.6}_{-3.1}$ \\
\vspace{0.5mm}
 12.0 & 1.7$^{+4.0}_{-1.4}$ \\
 \hline\hline \\  [0.5ex]  
 \end{tabular}
}
 \refstepcounter{table}\label{table:counts}
 \end{table}

In Figure~\ref{fig:counts} we also show the cumulative sub-mm
counts from the semi--analytic model {\sc galform} (Lacey et al.\ in prep.) constructed using the simulations of single-dish and ALMA
follow--up observations presented by \citet{Cowley15}. The current
version of the model adopts an Initial Mass Function (IMF) in
starbursts that is close to Salpeter, in contrast to previous versions
that required a top--heavy initial mass function to describe SMGs
(e.g.\ \citealt{Baugh05}). In this new model the intense starbursts in
SMGs are predominantly triggered by instabilities in gas--rich
discs. To ensure a fair comparison to the counts presented here, we
repeat the simulations presented in \citet{Cowley15} but adopt the
SCUBA--2 beam size and select sources with a flux density
$\ge$\,7.5\,mJy. As can be seen in Figure~\ref{fig:counts} the
predicted follow--up counts from the model are in broad agreement with
the counts presented here.

\subsection{Origin of Multiplicity}\label{sec:clustering}
A number of studies have investigated the environments of SMGs and
concluded that the population are strongly clustered
\citep{Blain04a,Scott06,Weiss09,Hickox12}, although studies have
questioned the robustness of these results \citep{Adelberger05,
  Williams11}. Similarly, a small number of single-dish sources have
been {\it resolved} into pairs of SMGs that have been
spectroscopically confirmed to lie at the same redshift
(e.g.\ \citealt{Tacconi06,Hodge13b,Ivison13}). Such small scale
over--densities of SMGs are unsurprising if these sources represent a
population of massive galaxies, potentially undergoing merger induced
star formation. Source multiplicity in SMGs thus offers one route to
investigate the environments of these sources on scales up to 140\,kpc
(the FWHM primary beam of ALMA at z\,$\sim$\,2.5). 

The origin of multiplicity in SMGs can be conclusively tested through
spectroscopic follow--up of the SMGs detected in each ALMA map. 
However, as we do not currently have spectroscopic redshifts for any of
the SMGs in our sample we instead use the number density of sources
fainter than the primary SMG in each ALMA map to assess their likely
association. If these secondary SMGs are simply line--of--sight
projections, then, in the absence of any bias, the number density of
sources should be equivalent to the background counts. However, we
must take into account the effect of blending on our initial sample
selection, which will enhance the number of SMGs with companions in
our target sample. Hence in the following analysis we {\it only} consider
ALMA maps in our sample that contain an SMG brighter than 8\,mJy. In
doing so we ensure that we only consider the maps that would have
been observed, regardless of whether blending with the detected
companion boosts the flux density of the single-dish source into our sample.       

There are eleven ALMA maps in our sample that contain an SMG
brighter than 8\,mJy, and we detect a total of twelve secondary SMGs in
these maps. To derive the surface density of these SMGs we must adopt
a flux limit for the sample. As the noise in the primary--beam
corrected maps increases with distance from the phase centre, and to
ensure that we have uniform coverage, we first remove one secondary
SMG with $S_{870}$\,$<$\,2\,mJy that would not have been detected at  
the edge of the ALMA primary beam. We calculate the area surveyed by
the ALMA primary beam in these maps and measure that the cumulative
number density of secondary SMGs brighter than 2\,mJy is
(5.5$_{-1.6}^{+2.2}$)\,$\times$\,10$ ^{4}$\,deg$^{-2}$ and we show this
point on Figure~\ref{fig:counts}. In these eleven maps we
expect to detect 0.14 SMGs at $S_{870}$\,$>$\,2\,mJy (adopting the
blank field counts in Figure~\ref{fig:counts}) but identify eleven
SMGs. Therefore, the number density of the secondary sources in our
maps is a factor of 80\,$\pm$\,30 times higher than the blank
field number counts, indicating that the brightest SMGs appear to
reside in over--dense regions. 

There is a small bias towards multiplicity in the
selection of our single-dish sources that arises from
observations of $>$\,8\,mJy SMGs that, due to random noise fluctuations,
scatter to lower values in the SCUBA--2 map. In such a scenario, an 8\,mJy
SMG is more likely to scatter back into our catalog if it
has a companion SMG that boosts the single-dish flux density above
our selection threshold. To determine the magnitude of this effect we
use a simulation of single-dish observations of SMGs, presented
by~\citet{Cowley15}. To remove any intrinsic clustering in the
simulation we randomise the positions of all of the input SMGs and
then apply the sample selection described above. The
resulting cumulative number density of secondary SMGs is a factor of
1.75\,$\pm$\,0.75\,times higher than the blank-field number
counts. While this analysis confirms that our sample has a small bias
due to noise, which increases the number of secondary SMGs, it is
clearly insufficient to explain the magnitude of the observed
offset.

Recently, theoretical predictions have been made for the origin of
multiplicity of single-dish submm sources (see
\citealt{Hayward13,Cowley15}). In the simulations presented by
\citet{Cowley15}, which are based on the semi-analytic model {\sc
  galform}, the majority of secondary SMGs are line-of-sight
projections, rather than physically associated sources. The apparent
over-density of secondary SMGs in our maps may indicate that the
brightest SMGs are more strongly clustered with fainter SMGs on
$\sim$\,arcsecond--scales than predicted by the model. 

Clearly, to conclusively confirm these physical associations requires
spectroscopic redshifts. We do not have spectroscopic redshifts for
any of the SMGs in our sample, and as shown by \citet{Simpson14} the
photometric redshifts of SMGs have considerable uncertainties, ruling
out the ability to perform this test with photometric redshifts
alone. Moreover, only 35\,\pc\, of the secondary SMGs we have
considered have a $K$--band counterpart (5\,$\sigma$ detection limit
25.0\,mag; see Ma et al. in prep.). The only way to conclusively test
if these sources are associated is through atomic or molecular
emission spectroscopy (i.e.\ [C{\sc ii}], $^{12}$\,CO) with sub-mm
interferometry (see \citealt{Weiss13}).

\section{Conclusion}
We have presented ALMA observations of 30 sub-mm bright
single-dish sources in the UDS field. These sources were selected
from 0.8\,deg$^2$, 850\,$\mu$m observations with SCUBA--2 at
the JCMT as part of the SCUBA--2 Cosmology Legacy Survey. The
main conclusions from our study are:

\begin{itemize}
\item The 30 ALMA maps in our sample have a resolution of
  0.35$''$\,$\times$\,0.25$''$, and median noise of
  0.21\,mJy\,beam$^{-1}$. Using tapered versions of these maps
  (median resolution 0.8$''$\,$\times$\,0.65$''$,
  $\sigma_{870}$\,=\,0.26\,mJy\,beam$^{-1}$) as detection images, we
  detect 52 SMGs at $>$\,4\,$\sigma$.  

%\item A comparison of the peak--to--integrated flux of the SMGs
%  demonstrates that the sources are resolved in both the
%  ``detection'' and ``high--resolution'' maps. As shown
%  in~\citet{Simpson15} the brightest SMGs in our sample have a median
%  intrinsic size of FWHM\,=\,$0.30$\,$\pm$\,0.04$''$, physical size of
%  (2.4\,$\pm$\,0.2)\,kpc, and we do not find any
%  evolution in the size of the rest-–frame far--infrared emission with either
%  redshift or 870\,$\mu$m flux density. 

\item We find that 61$^{+19}_{-15}$\,$\pc$\, of the single-dish sub-mm
  sources in our  sample are comprised of a blend of $\ge$\,2 SMGs
  brighter than $\gsim$\,1\,mJy (i.e.\ multiple ULIRGs). On average the
  brightest SMG in each ALMA map comprises 80$^{+6}_{-2}$\,\pc\, of
  the single-dish flux density, and where a secondary SMG is detected it
  contributes 25$^{+1}_{-5}$\,\pc\, to the total integrated flux
  density in the ALMA map. In two of our maps we do not detect any
  SMGs, and in ten maps we detect a single SMG. The remaining maps
  contain multiple SMGs, with 2, 3, or 4 SMGs detected in 14, 2 and 2
  maps, respectively. 

\item We compare our observations to the Cycle--0
  ALMA survey of single-dish sources, ALESS.  After
  accounting for the relative depths of both surveys we show that
  the fraction of sub-mm sources that are comprised 
  of a blend of multiple individual SMGs is consistent, at
  $\gsim$\,35\,\pc. However, in ALESS the brightest SMG in each ALMA
  map contains on average 65\,\pc\, of the single-dish flux density,
  compared to 80\,\pc\, for our sample. We show that this may be
  driven in part by the difference between the beam size of the initial
  single-dish selection for ALESS (LABOCA; beam-convolved FWHM\,=\,27.2$''$) and our survey
  (SCUBA--2; beam-convolved FWHM\,=\,20.5$''$). 

\item We construct the differential and cumulative sub-mm
  counts of SMGs from our ALMA observations. The multiplicity bias in
  single-dish sources means that the intrinsic cumulative number
  counts are 20\,\pc\, lower at $S_{870}$\,$>$\,7.5\,mJy than the
  single-dish SCUBA--2 survey, and 60\,\pc\, lower  at
  $S_{870}$\,$>$\,12\,mJy. We compare the counts derived from our
  survey to the theoretical models and demonstrate that the counts
  from the most recent {\sc galform} semi-analytic model \citep{Cowley15} are
  consistent with our results, at the flux density limit of our survey.   

\item The number density of secondary SMGs ($S_{870}$\,$>$\,2\,mJy)
  around the brightest sources in our sample is 80\,$\pm$\,30 times
  higher than expected from blank-field number counts. We caution that
  this result is still dominated by small number statistics, but we
  show that even after accounting for selection biases a significant fraction of these SMGs
  are likely to be physically associated. This suggests that the
  brightest SMGs reside in over--dense regions of SMGs

\end{itemize}

\section*{Acknowledgements}
We thank Adam Avison and the Manchester ALMA ARC node for their
assistance in verifying the calibration and imaging of our ALMA data. JMS
acknowledges the support of STFC studentship (ST/J501013/1). AMS
acknowledges financial support from an STFC Advanced Fellowship
(ST/H005234/1). IRS acknowledges support from the ERC Advanced
Investigator program DUSTYGAL 321334, an RS/Wolfson Merit Award and
STFC (ST/I001573/1). JEG acknowledges support from the Royal Society. RJI
acknowledges support from the European Research Council (ERC) in the
form of Advanced Grant, COSMICISM 321302. E. Ibar acknowledges funding
from CONICYT FONDECYT postdoctoral project N$^\circ$:3130504. KK
acknowledges support from the Swedish Research Council. JSD
acknowledges the support of the European Research Council via the
award of an Advanced Grant, and the contribution of the EC FP7 SPACE
project ASTRODEEP (312725). I. A acknowledges support from the grant CONACyT CB-2011-01-167291. 

This paper makes use of the following ALMA data:
ADS/JAO.ALMA$\#$2012.1.00090.S. ALMA is a partnership of ESO
(representing its member states), NSF (USA) and NINS (Japan), together
with NRC (Canada) and NSC and ASIAA (Taiwan), in cooperation with the
Republic of Chile. The Joint ALMA Observatory is operated by ESO,
AUI/NRAO and NAOJ. This publication also makes use of data taken with
the SCUBA--2 camera on the James Clerk Maxwell Telescope. The James
Clerk Maxwell Telescope is operated by the Joint Astronomy Centre on
behalf of the Science and Technology Facilities Council of the United
Kingdom, the National Research Council of Canada, and (until 31 March
2013) the Netherlands Organisation for Scientific Research. Additional
funds for the construction of SCUBA--2 were provided by the Canada
Foundation for Innovation. 

All data used in this analysis can be obtained from the ALMA archive.

\bibliographystyle{mn2e} 
\bibliography{ref.bib}

\end{document}